\documentclass[%
 aip,
 amsmath,amssymb,
 reprint,%
]{revtex4-1}

\usepackage{graphicx}
\usepackage{dcolumn}
\usepackage{bm}
\usepackage{color}
\usepackage[utf8]{inputenc}
\usepackage[T1]{fontenc}
\usepackage{mathptmx}

\begin{document}

\preprint{AIP/123-QED}

\title{Theory of exciton transport in molecular crystals strongly coupled to a cavity: A temperature-dependent variational approach}

\author{Jingyu Liu}
\author{Qing Zhao}
\author{Ning Wu}%
\email{wun1985@gmail.com}
\affiliation{Center for Quantum Technology Research, School of Physics, Beijing Institute of Technology, Beijing 100081, China}%

\date{\today}

\begin{abstract}
We present a semianalytical theory for exciton transport in organic molecular crystals interacting strongly with a single cavity mode. Based on the Holstein-Tavis-Cummings model and the Kubo formula, we derive an exciton mobility expression in the framework of a temperature-dependent variational canonical transformation, which can cover a wide range of exciton-vibration coupling, exciton-cavity coupling, and temperatures. A closed-form expression for the coherent part of the total mobility is obtained in the zeroth order of the exciton-vibration coupling, which demonstrates the significance of vibrationally dressed dark excitons in the determination of the transport mechanism. By performing numerical simulations on both the H- and J-aggregates, we find that the exciton-cavity coupling has significant effects on the total mobility: 1) At low temperatures, there exists an optimal exciton-cavity coupling strength for the H-aggregate at which a maximal mobility is reached, while the mobility in the J-aggregate decreases monotonically with increasing exciton-cavity coupling; 2) At high temperatures, the mobility in both types of aggregates get enhanced by the cavity. We illustrate the above-mentioned low-temperature optimal mobility observed in the H-aggregate by using realistic parameters at room temperature.
\end{abstract}

\maketitle

\section{\label{I}Introduction}
\par Exciton diffusion in molecular materials is a fundamental process relevant to a variety of physical phenomena, including organic semiconductors~\cite{Silbey2007}, solar cell physics~\cite{NatMat}, and excitation energy transfer in natural/artificial light-harvesting systems~\cite{Fleming2009,Scholes2011}, etc. In particular, the exciton transport in molecular crystals has been a long-studying topic dating back to the early theoretical investigations by Silbey and co-works in the 1970's~\cite{Silbey1971,Silbey1976,Silbey1976a,Silbey1977,Silbey1977a,Silbey1980,Munn1980}. In contrast to conventional inorganic crystals which can be described by band transport, phononic/vibrational degrees of freedom have to be included in the description of exciton and charge transport in organic crystals. The exciton transport properties in organic molecular crystals generally depend on a variety of factors, including the electronic transfer integrals between the excitons (or the excitonic bandwidth), the exciton-phonon coupling strength, the characteristic frequency of the phonons, and temperature, and so on.
\par Typical theoretical investigations on the charge-carrier transport in molecular crystals are usually based on the famous Holstein Hamiltonian~\cite{Hols}, and various theoretical methods were developed to treat the dynamics of the Holstein model (see Refs.~\cite{Silbey2007,Troisi2011} and references therein). Among these, the (variational) polaron transformation and its generalizations offers a promising approach to deal with the problem at a broad range of parameters~\cite{Silbey1976a,Silbey1977,Silbey1980,Silbey1985,Cheng2008,PRB2009,PRB2019}. In particular, Cheng and Silbey developed a finite-temperature variational approach by combining Merrifield's transformation with Bogoliubov's bound on the free energy of the composite system~\cite{Cheng2008}. Based on the polaron transformation, Ortmann \emph{et al}. presented a theoretical description of charge transport in molecular crystals and derived a mobility expression using the Kubo formula~\cite{PRB2009}.
\par Recently, the demonstration of strong coupling between confined light fields and organic matter has stimulated intensive interest in experimental/theoretical investigations of strong-coupling effects on charge-carrier transport~\cite{Orgiu2015,FJ2015,Genes2015,JYZhou2016,Genes2017,Zhong2017,JYZhou2018,FJ2018,JCP2019,PRE2019,Keeling2020}. Although in the framework the pure exciton model a dramatic enhancement of the exciton-type transport in organic materials has been revealed~\cite{Orgiu2015,FJ2015,Genes2015}, it is recognized that intramolecular vibrational mode should be included to obtain a more realistic description of the system~\cite{JCP2019,PRE2019,Keeling2020,Keeling2014,Spano2015,Spano2016,PRB2016,Spano2017,Spano2017a,Keeling2018,Jol2019}. The such obtained composite system involving excitonic, phononic, and photonic degrees of freedom is described by the so-called Holstein-Tavis-Cummings model, which was first termed by Herrera and Spano in Ref.~\cite{Spano2016}. In the framework of the Holstein-Tavis-Cummings Hamiltonian, Wu \emph{et al}.~\cite{PRB2016} generalized the Cheng-Silbey~\cite{Cheng2008} method to include the vibrational dressing of the cavity mode in the variational canonical transformation and provided a successful description of the static properties of the system. In special, a ground state involving all the excitonic, photonic, and vibrational degrees of freedom was demonstrated and named as a lower polaron polariton~\cite{PRB2016}.
\par In this work, we present a microscopic theory of exciton transport in molecular crystals strongly coupled to a cavity by extending the formalism developed in Ref.~\cite{PRB2016} to a time-dependent scenario. Based on the Holstein-Tavis-Cummings Hamiltonian and the Kubo formula, we derive an exciton mobility expression in the framework of the temperature-dependent variational canonical transformation~\cite{PRB2016} that can cover a wide range of exciton-vibration and exciton-cavity couplings, as well as temperatures. Due to the appearance the cavity mode, the mobility has three contributions, i.e., the conventional pure exciton part, the pure cavity part, and the cross term between the two. At low temperatures, it is found that the mobility in an H-aggregate depends nonmonotonically on the exciton-cavity coupling strength. That is, there exists an optimal exciton-cavity coupling strength at which the total mobility reaches a maximum. However, for the J-aggregate we observe a monotonic decrease of the total mobility with increasing exciton-cavity coupling in the low temperature regime. These observations are explained by obtaining an analytical expression for the coherent mobility (without vibration scattering), from which we identify that the interplay between vibrationally dressed dark excitons and the lower (upper) polaron polariton state determines the low-temperature exciton transport in the H-aggregates (J-aggregates). In the opposite limit with high temperatures, the mobility in both the two types of aggregates gets enhanced by the exciton-cavity coupling.
\par The rest of the paper is structured as follows. In Sec.~\ref{II}, we introduce the theoretical model and describe the generalized Merrifield varitional transformation in detail. In Sec.~\ref{III}, we derive the expressions for the total mobility and its coherent part. In Sec.~\ref{IV} we present numerical examples to illustrate the application of our formalism to the exciton mobility in the H- and J-aggregates. Conclusions are drawn in Sec.~\ref{V}.
\section{Model and methodology}\label{II}
\subsection{Hamiltonian}
\par
\begin{figure}
\includegraphics[width=.49\textwidth]{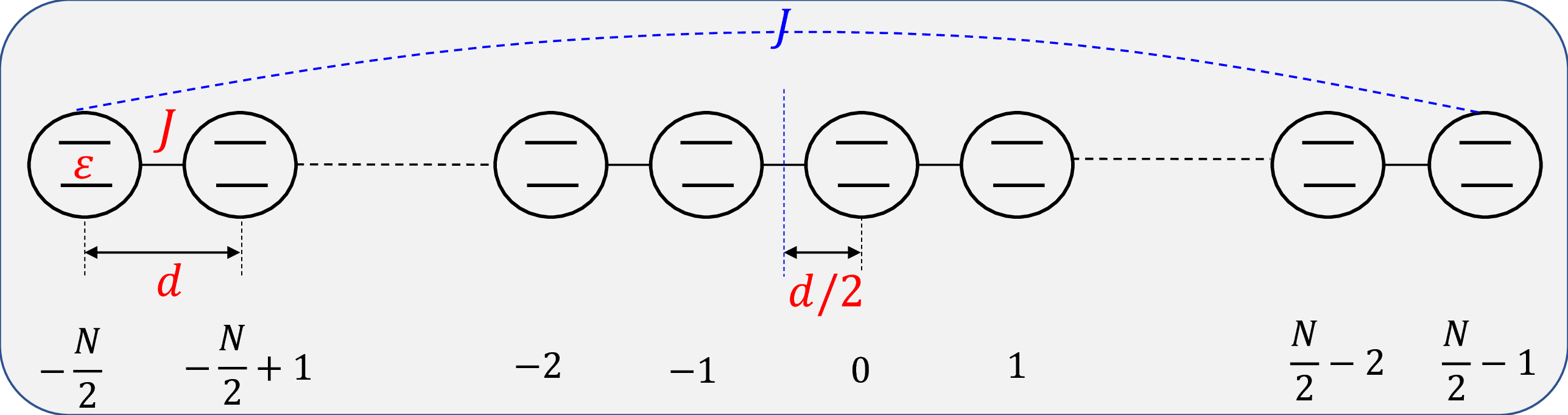}
\caption{A linear molecular chain consisting $N$ monomers is located in a single-mode cavity. The position of monomer $n$ is $R_n=d(2n+1)/2$ with $d$ the uniform lattice spacing.}
\label{Fig1}
\end{figure}

 We consider a one-dimensional molecular aggregate consisting of $N$ monomers located in a single-mode cavity. We assume $N$ is even for simplicity and the ensemble of monomers are arranged in a linear array with the origin of coordinates set at the middle point of the molecular chain, which is also the centroid of all the monomers in the chain. The reason for choosing the middle point of the molecular as the origin of coordinates will become clear below. The position of monomer $n$ ($n=-N/2,-N/2+1,\cdots,N/2-1$) is thus $R_n=d(2n+1)/2$, where $d$ is the uniform lattice spacing (see Fig.~\ref{Fig1}). By including the intramolecular vibrations, the system is described by the following Holstein-Tavis-Cummings Hamiltonian ($\hbar=1$)~\cite{Spano2015,Spano2016,PRB2016,Spano2017,Spano2017a,Keeling2018}
\begin{eqnarray}\label{H}
H&=&H_{\rm{mat}}+H_{\rm{c}}+H_{\rm{e-c}},\nonumber\\
H_{\rm{mat}}&=&H_{\rm{e}}+H_{\rm{v}}+H_{\rm{e-v}},\nonumber\\
H_{\rm{e}}&=&\sum^{N/2-1}_{n=-N/2}  \varepsilon a^\dag_na_n+J\sum^{N/2-2}_{n=-N/2}(a^\dag_na_{n+1}+a^\dag_{n+1}a_{n}),\nonumber\\
H_{\rm{v}}&=&\omega_0\sum^{N/2-1}_{n=-N/2} b^\dag_nb_n,\nonumber\\
 H_{\rm{e-v}}&=&\lambda  \omega_0\sum^{N/2-1}_{n=-N/2} a^\dag_na_n(b_n+b^\dag_n),\nonumber\\
H_{\rm{c}}&=&\omega_{\mathrm{c}}c^\dag c,~H_{\rm{e-c}}=\sum^{N/2-1}_{n=-N/2}g_n(a^\dag_n c+c^\dag a_n).
\end{eqnarray}
The material part $H_{\rm{mat}}$ is the one-dimensional Holstein model describing the molecular aggregate with intramolecular vibrations, where $J$ is the uniform nearest-neighbor electronic coupling. The creation operator $a^\dag_n$ ($b^\dag_n$) creates an exciton (vibration) on site $n$ with energy $\varepsilon$ ($\omega_0$). The linear exciton-vibration coupling is measured by the Huang-Rhys factor $\lambda^2$. Note that $H_{\rm{mat}}$ can be used to describe various organic systems including molecular crystals and organic semiconductors~\cite{Silbey2007,Silbey1980,Cheng2008}, J- and H-aggregates~\cite{JHSpano}, and light-harvesting compldex II~\cite{LHII1,LHII2}, etc. The single-mode cavity is described by $H_{\rm{c}}$, where $c^\dag$ creates a photon with frequency $\omega_{\rm{c}}$. $H_{\rm{e-c}}$ represents the exciton-cavity interaction with local strength $g_n\propto e^{-ikdn}$ (with $k$ the wavevector of the transverse standing wave in the cavity)~\cite{PRE2019}. The lattice spacing $d$ is usually in the nanometre range, so that $kd\ll 1$, we thus will use an approximated uniform coupling, $g_n=g$~\cite{Orgiu2015}, for not too large chains. We also employed the rotating wave approximation in $H_{\rm{e-c}}$ so that the total number of excitations $\sum_n a^\dag_n a_n+c^\dag c$ is conserved. This is a good approximation provided the ultrastrong-coupling regime is not reached~\cite{Spano2015,PRB2016,JCP2019}. Although the Holstein-Tavis-Cummings model is commonly used in the literature, we should keep in mind that it is derived under various approximations from first-principle Hamiltonians. For example, we have neglected the diamagnetic terms arising from the $\vec{A}^2$ term~\cite{Liberato}. As shown in Ref.~\cite{KeelingA2}, the diamagnetic terms can be formally removed by performing a Bogoliubov transformation of the cavity modes in the case of a multimode cavity, resulting in a renormalized exciton-cavity coupling.
\par For the sake of theoretical simplicity, we will allow for a boundary hopping of excitons between the two end monomers ($-N/2$ and $N/2-1$) to form a molecular ring with $a_{N/2}=a_{-N/2}$ (see the blue dashed curve Fig.~\ref{Fig1}). Such a choice of periodic boundary conditions allows us to work in the momentum space of the molecular ring, which results in only a single bright exciton for the pure exciton-photon system described by $H_{\mathrm{e}}+H_{\mathrm{c}}+H_{\mathrm{e-c}}$. Note that for a molecular chain with free ends there are more than one bright excitons due to a different form of the single-exciton dispersion~\cite{Spano1991,Wu2018}. We will work in the single-excitation subspace with $\sum_n a^\dag_n a_n + c^\dag c=1$, which allows us to truncate the number of cavity photons to be at most one. The creation operators under the above single-excitation approximation can be written as $a^\dag_n=|n\rangle\langle \rm{vac}|$ and $c^\dag=|c\rangle\langle \rm{vac}|$ with $|\rm{vac}\rangle$ the common vacuum of all the annihilation operators. By performing the following Fourier transform on the exciton operator,
 \begin{eqnarray}\label{FT}
a_n&=&\frac{1}{\sqrt{N}}\sum_k e^{ikdn}a_k,
\end{eqnarray}
where $k=-\frac{\pi}{d},-\frac{\pi}{d}+\frac{2\pi}{Nd},\cdots,\frac{\pi}{d}-\frac{2\pi}{Nd}$, the pure exciton-photon Hamiltonian can be written in the momentum space as
\begin{eqnarray}\label{P-crys}
&&H_{\mathrm{e}}+H_{\mathrm{c}}+H_{\mathrm{e-c}}\nonumber\\
&=&\sum_k(\varepsilon+2J\cos k)a^\dag_k a_k+\omega_{\mathrm{c}}c^\dag c+g\sqrt{N}(a^\dag_0 c+c^\dag a_0).\nonumber
\end{eqnarray}
We see that only the bright exciton state with zero momentum, $|k=0\rangle=a^\dag_{0}|\rm{vac}\rangle$, couples to the cavity field.
\subsection{The generalized Merrifield transformation}
\par To treat the exciton-vibration and exciton-cavity coupling at finite temperatures on an equal footing, we will employ a Merrifield transformation~\cite{Cheng2008,PRB2016} in which the variational parameters are determined by minimizing the Bogoliubov upper bound for the free energy of the whole system. It has been shown that the generalized Merrifield transformation could offer an accurate description of both static~\cite{PRB2016} and dynamical properties~\cite{Jol2019} for a wide range of exciton-vibration and exciton-cavity coupling, and temperatures.
\par Following Ref.~\cite{PRB2016}, we perform the following variational canonical transformation to $H$,
\begin{eqnarray}\label{MT}
\tilde{H}&=&e^{\mathcal{S}}He^{-\mathcal{S}},\nonumber\\
\mathcal{S}&=&-\sum_n a^\dag_n a_nB_n-c^\dag c B_{\rm{c}},
\end{eqnarray}
where
\begin{eqnarray}\label{Bn}
B_n&=&\sum_l f_l (b_{n+l}-b^\dag_{n+l}),~B_{\rm{c}}=h\sum_l (b_l-b^\dag_l)
\end{eqnarray}
are parameter-dependent vibrational operators. The variational parameters $\{f_l\}$ and $h$ can be chosen real~\cite{PRB2016} and will be determined in a self-consistent way by minimizing the free energy of the transformed Hamiltonian following Bogoliubov's inequality~\cite{feynman},
\begin{eqnarray}\label{BogoF}
F\leq F_0+\langle \mathcal{H}_1\rangle_{\mathcal{H}_0},
\end{eqnarray}
where $\mathcal{H}=\mathcal{H}_0+\mathcal{H}_1$ is a generic Hamiltonian and $F$ and $F_0$ are the respective free energies of $\mathcal{H}$ and $\mathcal{H}_0$, and $\langle \mathcal{H}_1\rangle_{\mathcal{H}_0}$ represents the thermal average of $\mathcal{H}_1$ over the canonical ensemble defined by $\mathcal{H}_0$.
\par Physically, the coefficient $f_l$ ($h$) measures the degree of dressing of the exciton at site $n$ (the cavity photon) by the vibrational mode on site $n+l$ (on each site). Note that it is necessary to take the vibrational dressing of the photon into account even if the cavity is not directly coupled to the vibrations~\cite{PRB2016}. We use a single dressing parameter $h$ for the photon since each monomer in the uniform molecular ring behaves equivalently to the photonic mode. Roughly speaking, we can imagine that the photonic excitation resides at the centroid, i.e, the centre of the ring, though it is unphysical to talk about the ``position" of a photon. This symmetrical consideration also leads us to choose the middle point of the original molecular open chain as the centre of coordinates for the monomers. The usual full polaron transformation in the context of the Holstein model corresponds to the case of $f_l=\delta_{l0}\lambda$ and $h=0$. By further noting that for a molecular ring the exciton-vibration system also holds mirror symmetry, $f_l=f_{N-l}$, the number of independent variational parameters is reduced to $\frac{N}{2}+1$, i.e., $\{f_0,f_1=f_{N-1},...,f_{\frac{N}{2}-1}=f_{\frac{N}{2}+1},f_{\frac{N}{2}}\}$ for even $N$~\cite{Cheng2008}.
\par The transformed Hamiltonian $\tilde{H}$ can be separated in a conventional way as
\begin{eqnarray}\label{Htilde}
\tilde{H}&=&\tilde{H}_{\mathrm{S}}+\tilde{V}+H_{\mathrm{v}}.
\end{eqnarray}
Here, the system part reads
\begin{eqnarray}\label{H_Stilde}
\tilde{H}_{\mathrm{S}}&=&\sum_{k\neq0}E_k a^\dag_ka_k+[E_0 a^\dag_0a_0+\tilde{g}\sqrt{N } (  a^\dag_0 c+ c^\dag a_0 )+\tilde{\omega}_{\mathrm{c}}c^\dag c],\nonumber\\
\end{eqnarray}
where
\begin{eqnarray}\label{bareband}
E_k=\varepsilon+\omega_0\left(\sum_m f^2_m-2\lambda f_0\right)+2\tilde{J}\cos kd
\end{eqnarray}
is the renormalized exciton dispersion. The renormalized parameters appearing in $\tilde{H}_{\mathrm{S}}$ are given by
\begin{eqnarray}\label{tilde_para}
\tilde{J}=J\Theta_1,~\tilde{g}=g\Theta,~\tilde{\omega}_{\mathrm{c}}=\omega_{\mathrm{c}}+Nh^2\omega_0,
\end{eqnarray}
with
\begin{eqnarray}\label{TT1}
\Theta&=&\langle e^{B_{\mathrm{c}}-B_n}\rangle_{\mathrm{v}}=e^{-\frac{1}{2}\coth\frac{\beta\omega_0}{2}\sum_l(f_l-h)^2},\nonumber\\
\Theta_{|n-n'|}&=&\langle e^{B_n-B_{n'}}\rangle_{\mathrm{v}}=e^{-\frac{1}{2}\coth\frac{\beta\omega_0}{2}\sum_l(f_{l-n}-f_{l-n'})^2},\nonumber\\
\end{eqnarray}
where $\langle ...\rangle_{\mathrm{v}}=\mathrm{Tr}_{\mathrm{v}}\{e^{-\beta H_{\mathrm{v}}}...\}/\mathrm{Tr}_{\mathrm{v}}\{e^{-\beta H_{\mathrm{v}}}\}$ (with $\beta=1/k_BT$ the inverse temperature) is the thermal average with respect to the vibrational modes. It can be seen from Eqs.~(\ref{tilde_para}) and (\ref{TT1}) that the presence of the exciton-vibration coupling decreases both the effective hopping integral $J$ and the exciton-cavity coupling $g$, but increases the effective cavity frequency $\omega_{\mathrm{c}}$.
\par The $N-1$ eigenstates of $\tilde{H}_{\mathrm{S}}$, $|k\rangle=a^\dag_k|\mathrm{vac}\rangle$ ($k\neq0$), are identical to the dark states of the pure exciton-photon Hamiltonian $H_{\mathrm{e}}+H_{\mathrm{c}}+H_{\mathrm{e-c}}$.  The interaction between the bright exciton $|k=0\rangle=a^\dag_0|\mathrm{vac}\rangle$ and the cavity mode leads to two branches of eigenmodes that diagonalize $\tilde{H}_{\mathrm{S}}$,
\begin{eqnarray}\label{Hsud}
\tilde{H}_{\mathrm{S}}&=&\sum_{k\neq0}E_k a^\dag_ka_k+E_{\mathrm{U}} a_{\mathrm{U}}^\dag a_{\mathrm{U}}+E_{\mathrm{D}} a_{\mathrm{D}}^\dag a_{\mathrm{D}},
\end{eqnarray}
where
\begin{eqnarray}\label{aUaD}
a_{\mathrm{U}}^\dag=Ca^\dag_0-Sc^\dag,~~a_{\mathrm{D}}^\dag=Sa^\dag_0+Cc^\dag
\end{eqnarray}
\begin{eqnarray}\label{Eud}
E_{\mathrm{U/D}}=\frac{E_0+\tilde{\omega}_{\mathrm{c}}}{2}\pm\sqrt{N\tilde{g}^2+\left(\frac{E_0-\tilde{\omega}_{\mathrm{c}}}{2}\right)^2},
\end{eqnarray}
with the mixing coefficients $C=\cos\frac{\theta}{2}$ and $S=\sin\frac{\theta}{2}$  determined by $\tan\theta=2\tilde{g}\sqrt{N}/(\tilde{\omega}_{\mathrm{c}}-E_0)$. As pointed out in Ref.~\cite{PRB2016}, although the $N+1$ states $\{|k\rangle|  k\neq 0\}  $ and $|U/D\rangle\equiv a^\dag_{\mathrm{U}/\mathrm{D}}|\mathrm{vac}\rangle$ respectively resemble the dark states and upper/lower exciton polariton states of the bare exciton-photon system, they actually live in the Merrifield frame and do not correspond to physical eigenstates. We follow Ref.~\cite{PRB2016} to call the quasiparticles associated with  $\{|k\rangle|  k\neq 0\}  $ and $|U/D\rangle$ Merrifield dark excitons and Merrifield upper/lower polaritons, respectively. By applying the unitary transformation $e^{-\mathcal{S}}$ to the Merrifeld states, we obtain the $N+1$ physical states in the original frame, $\{e^{-\mathcal{S}}|k\rangle|k\neq 0\} $ and $e^{-\mathcal{S}}|U/D\rangle$, which are approximate eigenstates (to the zeroth order of the residue interaction $\tilde{V}$) of $H$. Due to the complex structure of the generator $\mathcal{S}$, the $N-1$ states $\{e^{-\mathcal{S}}|k\rangle\}$ generated by the Merrifield dark excitons are actually mixtures of vibrationally dressed dark and bright states. The state $e^{-\mathcal{S}}|U\rangle$ ($e^{-\mathcal{S}}|D\rangle$) is a mixture of vibrationally dressed dark and bright excitonic, as well as photonic states, which is referred to as an upper (lower) polaron polarition state~\cite{PRB2016}. Note that states with similar structures to the polaron polaritons are also revealed in Refs.~\cite{Spano2017,Spano2017a} in the case of $J=0$.
\par The explicit form of the residue interaction $\tilde{V}$ can be found in Ref.~\cite{PRB2016}. By construction, the thermal average of $\tilde{V}$ vanishes, $\langle\tilde{V}\rangle_{\mathrm{v}}=0$. By setting $\mathcal{H}_0$ ($\mathcal{H}_1$) to be $\tilde{H}_{\mathrm{S}}+H_{\mathrm{v}}$ ($\tilde{V}$) in Eq.~(\ref{BogoF}), we are now ready to minimize the Bogoliubov upper bound for the free energy of $\tilde{H}$:
\begin{eqnarray}\label{BB}
F_{\mathrm{B}}=-\frac{1}{\beta}\ln \mathrm{Tr} e^{-\beta(\tilde{H}_{\mathrm{S}}+H_{\mathrm{v}})}=-\frac{1}{\beta}\ln Z_{\mathrm{S}}-\frac{1}{\beta}\ln Z_{\mathrm{v}},
\end{eqnarray}
where $Z_{\mathrm{S}}=\sum_{\eta=\{ k(\neq0),\mathrm{U,D}\}} e^{-\beta E_\eta}$ is the partition function for $\tilde{H}_{\mathrm{S}}$, and $Z_{\mathrm{v}}$ is the partition function of the free vibrational modes. Since $Z_{\mathrm{v}}$ does not depend on the variational parameters, we only need to minimize $-\frac{1}{\beta}\ln Z_{\mathrm{S}}$ in Eq.~(\ref{BB}), which results in the saddle-point conditions $\{\partial Z_{\mathrm{S}}/\partial f_n=0\}$ and $\partial Z_{\mathrm{S}}/\partial h=0$ that need to be solved self-consistently (see Ref.~\cite{PRB2016} for the explicit forms of the saddle-point equations). Note that the such obtained $F_{\mathrm{B}}$ gives an upper bound for the true free energy of the whole system.
\section{Exciton mobility}\label{III}
\par Strictly speaking, the connection between exciton transport and charge transport is not obvious as excitons are in principle neutral quasiparticles. Nevertheless, Munn and Silbey argued that the charge-carrier drift mobility and exciton diffusion coefficient are proportional to each other~\cite{Silbey1980}. Moreover, the diffusion coefficient can be obtained by either calculating the time derivative of the mean-square displacement of an exciton in the long-time limit, or by using the Kubo formalism~\cite{Silbey1980}, which will be employed below to evaluated the exciton mobility.
\subsection{Basic formulas}
\par To obtain an expression for the exciton current, we consider the following ``position operator" for an exciton
\begin{eqnarray}\label{X}
X=\sum^{N/2-1}_{n=-N/2}R_n a^\dag_n a_n,
\end{eqnarray}
which acts on the local exciton state $|n\rangle$ giving
\begin{eqnarray}\label{Xn}
X|n\rangle=R_n|n\rangle.
\end{eqnarray}
The operator $X$ is an analog of the polarization operator for electric charges~\cite{PRB2009}. We define the exciton current (the ``velocity") operator $\mathcal{J}$ as the time derivative of $X$:
\begin{eqnarray}\label{current}
\mathcal{J}&=&\frac{dX}{dt}=-i[X,H] =\mathcal{J}_a+\mathcal{J}_{\mathrm{c}},
\end{eqnarray}
where
\begin{eqnarray}\label{currenta}
\mathcal{J}_a&=& i d J\sum_n( a^\dag_{n}a_{n+1}- a^\dag_{n+1}a_{n}),\nonumber\\
\mathcal{J}_{\mathrm{c}}&=& -i  g\sum_nR_n(a^\dag_n c-c^\dag a_n).
\end{eqnarray}
The two terms in Eq.~(\ref{current}) represent different contributions to the current operator: $\mathcal{J}_a$ corresponds to the usual exciton hopping between nearest-neighboring sites, while $\mathcal{J}_{\mathrm{c}}$ accounts for the exchange between the excitonic and the photonic excitations.
\par The mobility along the molecular chain can be obtained by the Kubo formula~\cite{Mahan}
\begin{eqnarray}\label{mob}
\mu=\frac{\beta}{2N_e} \int^\infty_{-\infty}dt\langle \mathcal{J}(t)\mathcal{J}(0)\rangle_{H},
\end{eqnarray}
where $\mathcal{J}(t)=e^{iHt}\mathcal{J}e^{-iHt}$, $N_e$ is the total number of excitons in the system, and the average is defined as the thermal average with respect to the full Hamiltonian $H$ as $\langle\cdots\rangle_{H}=\mathrm{Tr}(\cdots e^{-\beta H })/\mathrm{Tr}(e^{-\beta H })$. Using Eq.~(\ref{MT}), the current-current correlation function can be reexpressed in the Merrifield frame as
\begin{eqnarray}\label{cu-cu}
 &&\langle \mathcal{J}(t)\mathcal{J}(0)\rangle_{H}=\langle \tilde{\mathcal{J}}(t)\tilde{\mathcal{J}}(0) \rangle_{\tilde{H}}\nonumber\\
 &=&\langle \tilde{\mathcal{J}}_a(t)\tilde{\mathcal{J}}_a(0) \rangle_{\tilde{H}}+\langle \tilde{\mathcal{J}}_a(t)\tilde{\mathcal{J}}_{\mathrm{c}}(0) \rangle_{\tilde{H}}\nonumber\\
 &&+\langle \tilde{\mathcal{J}}_{\mathrm{c}}(t)\tilde{\mathcal{J}}_a(0) \rangle_{\tilde{H}}+\langle \tilde{\mathcal{J}}_{\mathrm{c}}(t)\tilde{\mathcal{J}}_{\mathrm{c}}(0) \rangle_{\tilde{H}}.
\end{eqnarray}
Here, the transformed current operators in the Merrifield frame and in the Heisenberg picture read
\begin{eqnarray}\label{Ja}
\tilde{\mathcal{J}}_a(t)&=& e^{i\tilde{H}t}\tilde{\mathcal{J}}_a e^{-i\tilde{H}t}\nonumber\\
&=&  i d J\sum_n[a^\dag_{n}(t)a_{n+1}(t)e^{B_{n+1}(t)-B_n(t)}\nonumber\\
&&-e^{B_{n}(t)-B_{n+1}(t) }a^\dag_{n+1}(t)a_{n}(t)],
\end{eqnarray}
and
\begin{eqnarray}\label{Jc}
\tilde{\mathcal{J}}_{\mathrm{c}}(t)&=& e^{i\tilde{H}t}\tilde{\mathcal{J}}_{\mathrm{c}} e^{-i\tilde{H}t}\nonumber\\
&=&-i  g\sum_nR_n[a^\dag_n(t) c(t) e^{B_{\mathrm{c}}(t)-B_n(t)}\nonumber\\
&&-e^{B_n(t)-B_{\mathrm{c}}(t)}c^\dag(t) a_n(t)].
\end{eqnarray}
\par The calculation of the Heisenberg picture operators appearing in the above equations seems challenging due to the complicated form of $\tilde{H}$. To proceed we have to resort to approximations. To this end, we note that the zeroth order energy $E_{\mathrm{D}}$ of $\tilde{H}_{\mathrm{S}}$ provides a good approximation for the true ground state energy of $\tilde{H}$~\cite{PRB2016} at zero temperature. We thus adopt the zero-order Hamiltonian
\begin{eqnarray}\label{HH0}
\tilde{H}\approx\tilde{H}_0=\tilde{H}_{\mathrm{S}}+H_{\mathrm{v}},
\end{eqnarray}
in the calculation of the current operators. This zeroth-order approximation is also employed in Ref.~\cite{PRB2009} to study charge transport in organic crystals in the framework the full polaron transformation. Under the approximation given by Eq.~(\ref{HH0}), the Heisenberg picture operators appearing in Eqs.~(\ref{Ja}) and (\ref{Jc}) can be calculated as
\begin{eqnarray}\label{Bjt}
B_n(t)&=&\sum_l f_l(b_{n+l}e^{-i\omega_0t}-b^\dag_{n+l}e^{i\omega_0t}),\nonumber\\
B_{\mathrm{c}}(t)&=&h\sum_l(b_le^{-i\omega_0t}-b^\dag_le^{i\omega_0t}),
\end{eqnarray}
and
\begin{eqnarray}\label{ajt}
a_n(t)&=& \frac{1}{\sqrt{N}}\sum^{N+1}_{\eta=1} e^{iK_\eta nd-i\mathcal{E}_\eta t}x_\eta f_\eta,\nonumber\\
c(t)&=&\sum^{N+1}_{\eta=1} e^{-i\mathcal{E}_\eta t}y_\eta f_\eta,
\end{eqnarray}
where we introduced (for $\eta=1,2,\cdots, N+1$)
\begin{eqnarray}
x_\eta&=&\{1,\cdots,1,C,S\}\nonumber\\
y_\eta&=&\{0,\cdots,0,-S,C\},\nonumber\\
K_\eta&=&\{k(\neq0),0,0\},\nonumber\\
\mathcal{E}_\eta&=&\{E_{k(\neq0)},E_{\mathrm{U}},E_{\mathrm{D}}\},\nonumber\\
f_\eta&=&\{a_{k(\neq0)},a_{\mathrm{U}},a_{\mathrm{D}}\}.\nonumber
\end{eqnarray}
In Appendix~\ref{AppA}, we list the explicit expressions for the four terms in the current-current correlation function given by Eq.~(\ref{cu-cu}). As an example, we just mention the cross term
\begin{widetext}
\begin{eqnarray}
 &&\langle\tilde{\mathcal{J}}_a(t)  \tilde{\mathcal{J}}_{\mathrm{c}}(0) \rangle_{\tilde{H}}\nonumber\\
  &=&dJg\sum_{jj'}R_{j'}\Big[\langle a^\dag_{j}(t)a_{j+1}(t) a^\dag_{j'}c\rangle_{\tilde{H}_{\mathrm{S}}}\langle e^{B_{j+1}(t)-B_j(t)}e^{B_{\mathrm{c}}-B_{j'}} \rangle_{H_{\mathrm{v}}}-  \langle a^\dag_{j}(t)a_{j+1}(t) c^\dag a_{j'}\rangle_{\tilde{H}_{\mathrm{S}}}\langle e^{B_{j+1}(t)-B_j(t)}e^{B_{j'}-B_{\mathrm{c}}} \rangle_{H_{\mathrm{v}}}\nonumber\\
& &-  \langle a^\dag_{j+1}(t)a_{j}(t) a^\dag_{j'}c\rangle_{\tilde{H}_{\mathrm{S}}}\langle e^{B_{j}(t)-B_{j+1}(t)}e^{B_{\mathrm{c}}-B_{j'}} \rangle_{H_{\mathrm{v}}}+   \langle a^\dag_{j+1}(t)a_{j}(t) c^\dag a_{j'}\rangle_{\tilde{H}_{\mathrm{S}}}\langle e^{B_{j}(t)-B_{j+1}(t)}e^{B_{j'}-B_{\mathrm{c}}} \rangle_{H_{\mathrm{v}}}\Big].
\end{eqnarray}
\end{widetext}
Typical thermal averages appearing in the above equation can be obtained as, e.g., (see Appendix~\ref{AppA} for the derivation)
\begin{eqnarray}
\langle a^\dag_{r}(t)a_{l}(t) a^\dag_{m}c\rangle_{\tilde{H}_{\mathrm{S}}}&=&\frac{1}{Z_{\mathrm{S}} N\sqrt{N}}\sum_{\eta} x_\eta y_\eta e^{-\beta\mathcal{E}_\eta-iK_\eta rd+i\mathcal{E}_\eta t}\nonumber\\
&&\sum_{\eta'} x^2_{\eta'} e^{iK_{\eta'} (l-m)d-i\mathcal{E}_{\eta'} t},
\end{eqnarray}
and
\begin{eqnarray}\label{eBeB}
&&\langle e^{B_{l}(t)-B_r(t)}e^{B_{\mathrm{c}}-B_{m}} \rangle_{H_{\mathrm{v}}}\nonumber\\
& =&\Theta\Theta_1 e^{-\Phi_{\omega_0}(t)\sum_s (f_{s-l}-f_{s-r})(h-f_{s-m}) },
\end{eqnarray}
where $\Phi_{\omega_0}(t)=n_{\omega_0}e^{i\omega_0t}+(1+n_{\omega_0})e^{-i\omega_0t}$ with $n_{\omega_0}=1/(e^{\beta\omega_0}-1)$ the Bose-Einstein distribution function. The two factors $\Theta$ and $\Theta_1$ are given by Eq.~(\ref{TT1}) and give rise to the band narrowing effect, while the exponential factor in Eq.~(\ref{eBeB}) describes the vibration scattering events.
\subsection{Coherent contribution}
\par  Following Ref.~\cite{PRB2009}, we separate the contributions to the mobility into a coherent part (without vibration scattering) and an incoherent part (scattering processes by the vibrations),
\begin{eqnarray}
\mu=\mu^{(\mathrm{coh})}+\mu^{(\mathrm{inc})},
\end{eqnarray}
which is achieved by, e.g., splitting the exponential factor in Eq.~(\ref{eBeB}) according to $e^{-\Phi_{\omega_0}(t)\sum_s (f_{s-l}-f_{s-r})(h-f_{s-m}) }=1+[e^{-\Phi_{\omega_0}(t)\sum_s (f_{s-l}-f_{s-r})(h-f_{s-m}) }-1]$, where the unity corresponds to coherent transport in the zeroth order of exciton-vibration coupling and the second term corresponds to incoherent transport~\cite{PRB2009}. In Appendix~\ref{AppB} we show that only the first and the last term in the coherent contribution to the current-current correlation function, $\langle \tilde{\mathcal{J}}_a(t)\tilde{\mathcal{J}}_a(0) \rangle^{\mathrm{coh}}_{\tilde{H} }$ and $\langle \tilde{\mathcal{J}}_{\mathrm{c}}(t)\tilde{\mathcal{J}}_{\mathrm{c}}(0) \rangle^{\mathrm{coh}}_{\tilde{H} }$, survive and lead to
\begin{eqnarray}\label{mucoh}
\mu^{(\mathrm{coh})}&=&\frac{\beta}{2 N_e} \int^\infty_{-\infty}dt[\langle \tilde{\mathcal{J}}_a(t)\tilde{\mathcal{J}}_a  \rangle^{\mathrm{coh}}_{\tilde{H}}+\langle \tilde{\mathcal{J}}_{\mathrm{c}}(t)\tilde{\mathcal{J}}_{\mathrm{c}}  \rangle^{\mathrm{coh}}_{\tilde{H}}]\nonumber\\
&=& \mu^{(\mathrm{coh})}_{a}+\mu^{(\mathrm{coh})}_{\mathrm{c}},
\end{eqnarray}
where the two contributions read
\begin{eqnarray}\label{mua}
\mu^{(\mathrm{coh})}_a&=&  \frac{2\beta  d^2}{  N_e Z_{\mathrm{S}}} \tilde{J}^2 \int^\infty_{-\infty}dt \sum^{N-1}_{\eta=1 }e^{-\beta \mathcal{E}_\eta}    (1-\cos^2K_\eta d),\nonumber\\
\end{eqnarray}
and
\begin{widetext}
\begin{eqnarray}\label{muc}
\mu^{(\mathrm{coh})}_{\mathrm{c}}&=& \frac{\beta  d^2(\tilde{g}\sqrt{N})^2 }{4 N_e Z_{\mathrm{S}}}\times\nonumber\\
&& \int^\infty_{-\infty}dt\sum^{N-1}_{\eta=1}   \frac{ S^2[e^{-\beta\mathcal{E}_\eta}e^{i(\mathcal{E}_\eta-E_{\mathrm{U}})t}+e^{-\beta E_{\mathrm{U}}}e^{-i(\mathcal{E}_\eta-E_{\mathrm{U}})t}]  +C^2[e^{-\beta\mathcal{E}_\eta}e^{i(\mathcal{E}_\eta-E_{\mathrm{D}})t}+e^{-\beta E_{\mathrm{D}}}e^{-i(\mathcal{E}_\eta-E_{\mathrm{D}})t}]  }{ 1-\cos K_\eta d }.
\end{eqnarray}
It is apparent that $\mu^{(\mathrm{coh})}_{\mathrm{c}}$ results from the exciton-cavity coupling and the time integration yields the Dirac delta function, $\delta(\mathcal{E}_\eta-E_{\mathrm{U/D}})$, which means that only those Merrifield dark states $|K_\eta\rangle$ resonant with the two Merrifield polaritons have finite contributions to the coherent mobility. Actually, by introducing the density of states of the Merrifield dark excitons
\begin{eqnarray}\label{DOS}
\rho(\epsilon)\equiv \sum^{N-1}_{\eta=1}\delta(\epsilon-\mathcal{E}_\eta),
\end{eqnarray}
we can rewrite $\mu^{(\mathrm{coh})}_{\mathrm{c}}$ as
\begin{eqnarray}\label{muc1}
\mu^{(\mathrm{coh})}_{\mathrm{c}}&=& \frac{2\beta  d^2(\tilde{g}\sqrt{N})^2 \pi\tilde{J}  }{ N_e Z_{\mathrm{S}}} \left[\frac{\rho (E_{\mathrm{U}})S^2  e^{-\beta E_{\mathrm{U}}}     }{ 2\tilde{J}-E_{\mathrm{U}}+\varepsilon+\omega_0\left(\sum_m f^2_m-2\lambda f_0\right) } + \frac{\rho (E_{\mathrm{D}}) C^2   e^{-\beta E_{\mathrm{D}}}   }{ 2\tilde{J}-E_{\mathrm{D}}+\varepsilon+\omega_0\left(\sum_m f^2_m-2\lambda f_0\right) }\right],
\end{eqnarray}
\end{widetext}
where we have used Eq.~(\ref{bareband}). As $N\to\infty$, the Merrifield dark excitons form a quasi-continuous band and $\rho(\epsilon)$ tends to be a continuous function~\cite{Sutton}
\begin{eqnarray}
\rho(\epsilon)&=& \frac{N }{\pi}\frac{1}{  \sqrt{4\tilde{J}^2-[\epsilon-\varepsilon-\omega_0\left(\sum_m f^2_m-2\lambda f_0\right)]^2}}.\nonumber\\
\end{eqnarray}
It is clearly seen from Eq.~(\ref{muc1}) that $\mu^{(\mathrm{coh})}_{\mathrm{c}}$ is finite only if the Merrifield upper or lower polariton lies within the Merrifield dark excitonic band~\cite{PRB2006}.
\par Let us turn now to $\mu^{(\mathrm{coh})}_a$, which diverges since exciton scattering by vibrations is absent in the zeroth order of the exciton-vibration coupling. Nevertheless, in real materials other types of scattering mechanism such as impurities or static disorders may exist and lead to a reduction of the mean free path of the excitons. To account for such processes, we follow Refs.~\cite{PRB2009,PRX} to introduce a Gaussian damping in the integration over time, i.e., $\int dt\to \int dt e^{-(t/\tau)^2}$, where $\tau$ is a finite scattering time. In this way, the coherent mobility can be obtained as
\begin{widetext}
\begin{eqnarray}\label{mob-coh}
\mu^{(\mathrm{coh})}&=& \frac{\sqrt{\pi}\tau\beta  ( d\tilde{J})^2}{  N_e Z_{\mathrm{S}}\hbar^2}    \sum^{N-1}_{\eta=1 }e^{-\beta E_\eta}    (1-\cos 2K_\eta) \nonumber\\
&&+\frac{\sqrt{\pi}\tau\beta  (d\tilde{g})^2 N}{4 N_e Z_{\mathrm{S}}\hbar^2}  \sum^{N-1}_{\eta=1}[S^2(e^{-\beta\mathcal{E}_\eta} +e^{-\beta E_{\mathrm{U}}})  e^{-\frac{1}{4}\tau^2(\mathcal{E}_\eta-E_{\mathrm{U}})^2} +C^2(e^{-\beta\mathcal{E}_\eta} +e^{-\beta E_{\mathrm{D}}})  e^{-\frac{1}{4}\tau^2(\mathcal{E}_\eta-E_{\mathrm{D}})^2} ]  \frac{ 1 }{ 1-\cos K_\eta d },\nonumber\\
\end{eqnarray}
\end{widetext}
where we restored $\hbar$. From the relation $c^\dag=Ca^\dag_{\mathrm{D}}-Sa^\dag_{\mathrm{U}}$, we expect that the term proportional to $S^2$ ($C^2$) in the second line of Eq.~(\ref{mob-coh}) corresponds to the contribution from the upper (lower) polaron polariton, which will be confirmed in the numerical simulations performed below. The coherent mobility $\mu^{(\mathrm{coh})}$ mainly determines the low-temperature behavior of the total mobility.
\section{Numerical results}\label{IV}
\subsection{Coherent mobility}
\begin{figure}
\includegraphics[width=.48\textwidth]{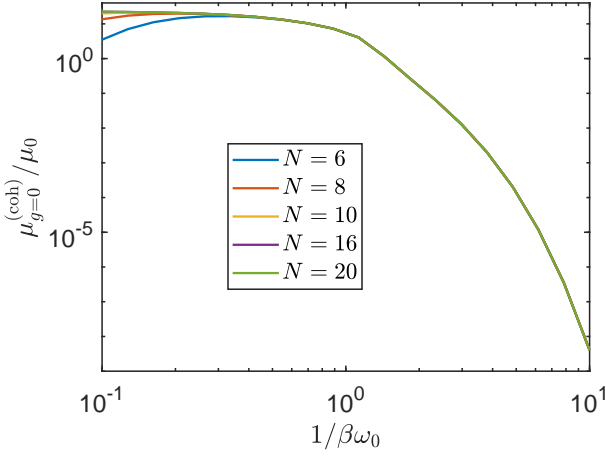}
\caption{Reduced coherent mobility $\mu^{(\mathrm{coh})}_{g=0}/\mu_0$ of an H-aggregate as a function of the reduced temperature $1/\beta\omega_0$ in the absence of the exciton-cavity coupling. Parameters: $g=0$, $\varepsilon/\omega_0=11.8$, $J/\omega_0=0.5$, $\omega_{\mathrm{c}}/\omega_0=11$, $\tau\omega_0=30$, and $\lambda=0.7$. The unit of the mobility is $\mu_0=d^2/(N_e\hbar)$.}
\label{Fig2}
\end{figure}
\par In order to demonstrate the effect of exciton-cavity interaction, we first consider the coherent mobility for $g=0$. In this case it can be seen from Eq.~(\ref{mob-coh}) that $\mu^{(\mathrm{coh})}_{g=0}$ is proportional to the truncation time $\tau$ and depend heavily on the structure of the energy levels, $\{E_k\}$,
\begin{eqnarray}\label{mob-cohg0}
\mu^{(\mathrm{coh})}_{g=0}&=&   \frac{\sqrt{\pi}\tau\beta  ( d\tilde{J})^2}{  N_e\hbar^2 }  \left(1-  \sum_{k }\frac{e^{-\beta E_k}}{Z_{\mathrm{S}}} \cos 2k\right).
\end{eqnarray}
It is well known that the exciton-vibration coupling will give rise to the so-called band narrowing with increasing temperature. In the high-temperature limit, $\mu^{(\mathrm{coh})}_{g=0}$ behaves like
\begin{eqnarray}\label{mob-cohg0highT}
\mu^{(\mathrm{coh})}_{g=0}&\to&  \frac{\sqrt{\pi}\tau   ( d\tilde{J})^2}{  N_e k_BT\hbar^2 },~\mathrm{as}~T\to\infty,
\end{eqnarray}
since $E_k$ becomes dispersionless. In Fig.~\ref{Fig2} we plot $\mu^{(\mathrm{coh})}_{g=0}$ [in unit of $\mu_0\equiv d^2/(N_e\hbar)$] as a function of the dimensionless temperature $1/\beta\omega_0$  for various numbers of monomers. Note that $\mu^{(\mathrm{coh})}_{g=0}$ is a nonmonotonic function of $T$ for short chains with $N\leq 8$. The suppression of the coherent mobility at low temperatures can be understood from the fact that the energy gap between different energy levels is too large to provide efficient transport channels due to the low thermal excitations of excitonic states. Note that convergent results are observed for $N\ge16$.
\par We now turn to discuss the effects of finite exciton-cavity interaction on the coherent mobility in both the H-aggregate ($J>0$) and J-aggregate ($J<0$). It is known that the absorption spectrum of an H-aggregate (J-aggregate) exhibit blue (red) shift relative to the monomer excitation energy in the absence of the exciton-cavity coupling~\cite{ACR2010,ARPC2014}.
\begin{figure}
\includegraphics[width=.52\textwidth]{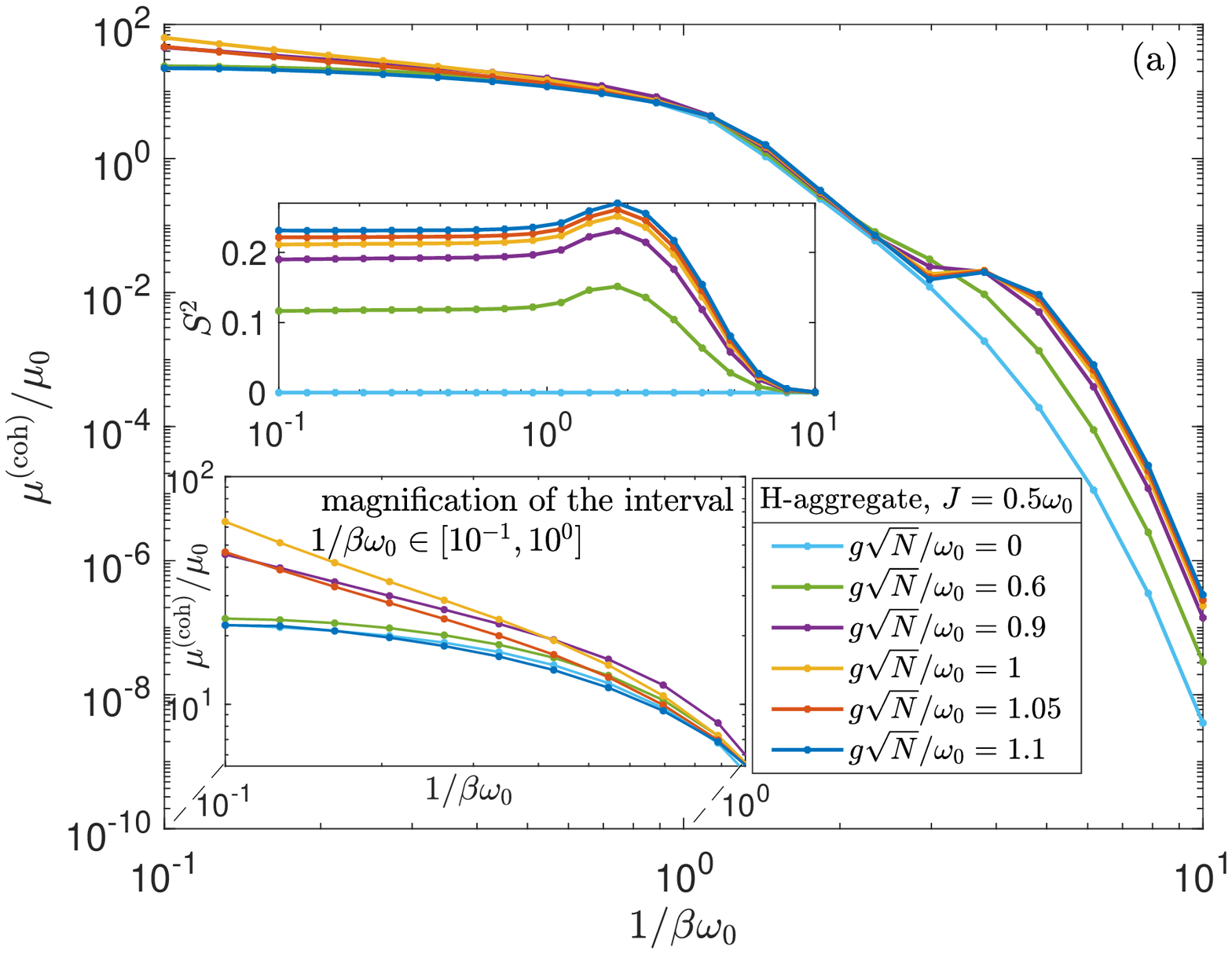}
\includegraphics[width=.52\textwidth]{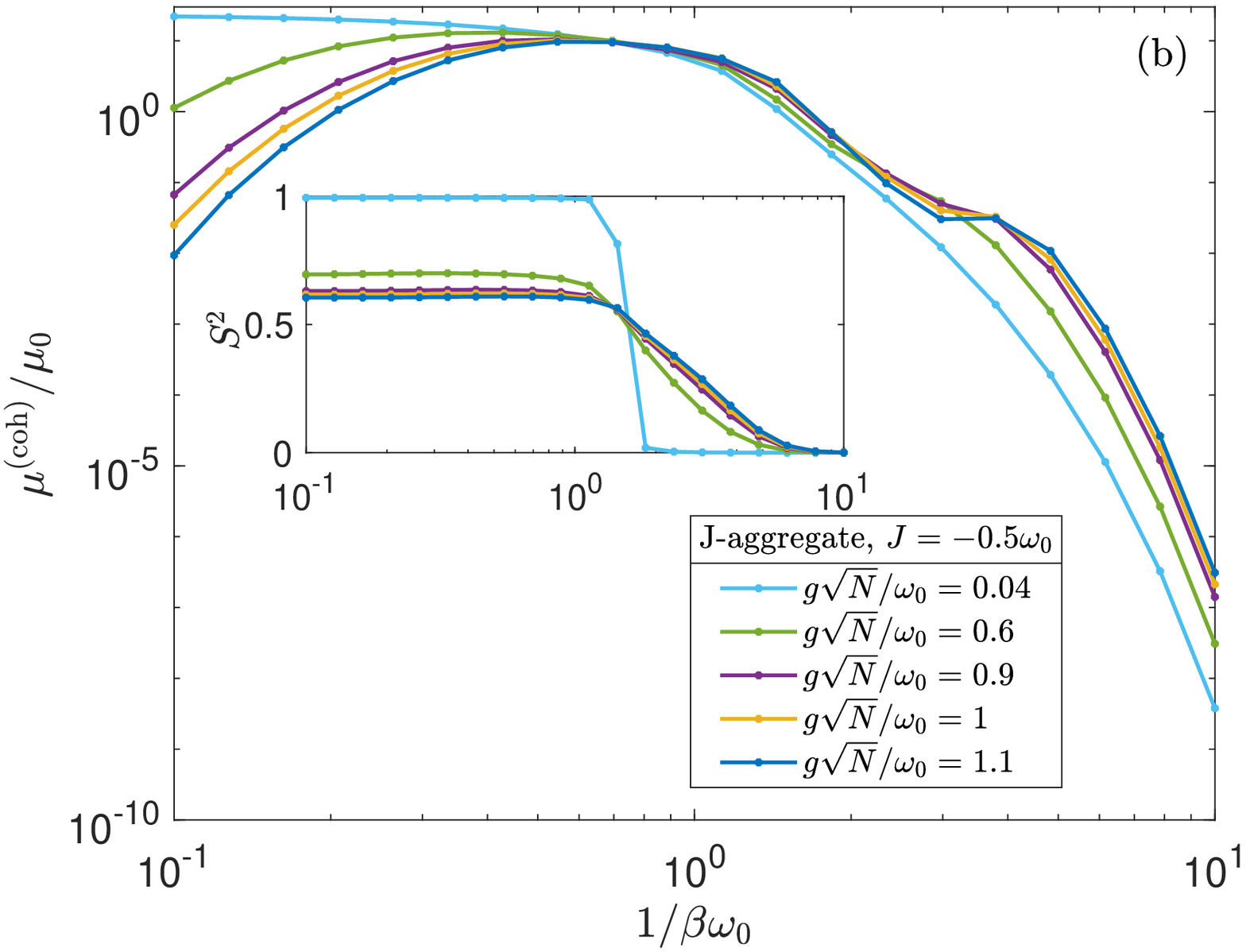}
\caption{Reduced coherent mobility of (a) an H-aggregate with $J/\omega_0=0.5$; (b) a J-aggregate with $J/\omega_0=-0.5$ for various values of $g\sqrt{N}/\omega_0$. The upper insets in the two panels show the corresponding variation of the mixing coefficient $S^2$ [see Eq.~(\ref{aUaD})]. Other parameters: $N=16$, $\tau\omega_0=30$, $\varepsilon/\omega_0=11.8$, $\omega_{\mathrm{c}}/\omega_0=11$, and $\lambda=0.7$. }
\label{Fig3}
\end{figure}
\par
\begin{figure}
\includegraphics[width=.52\textwidth]{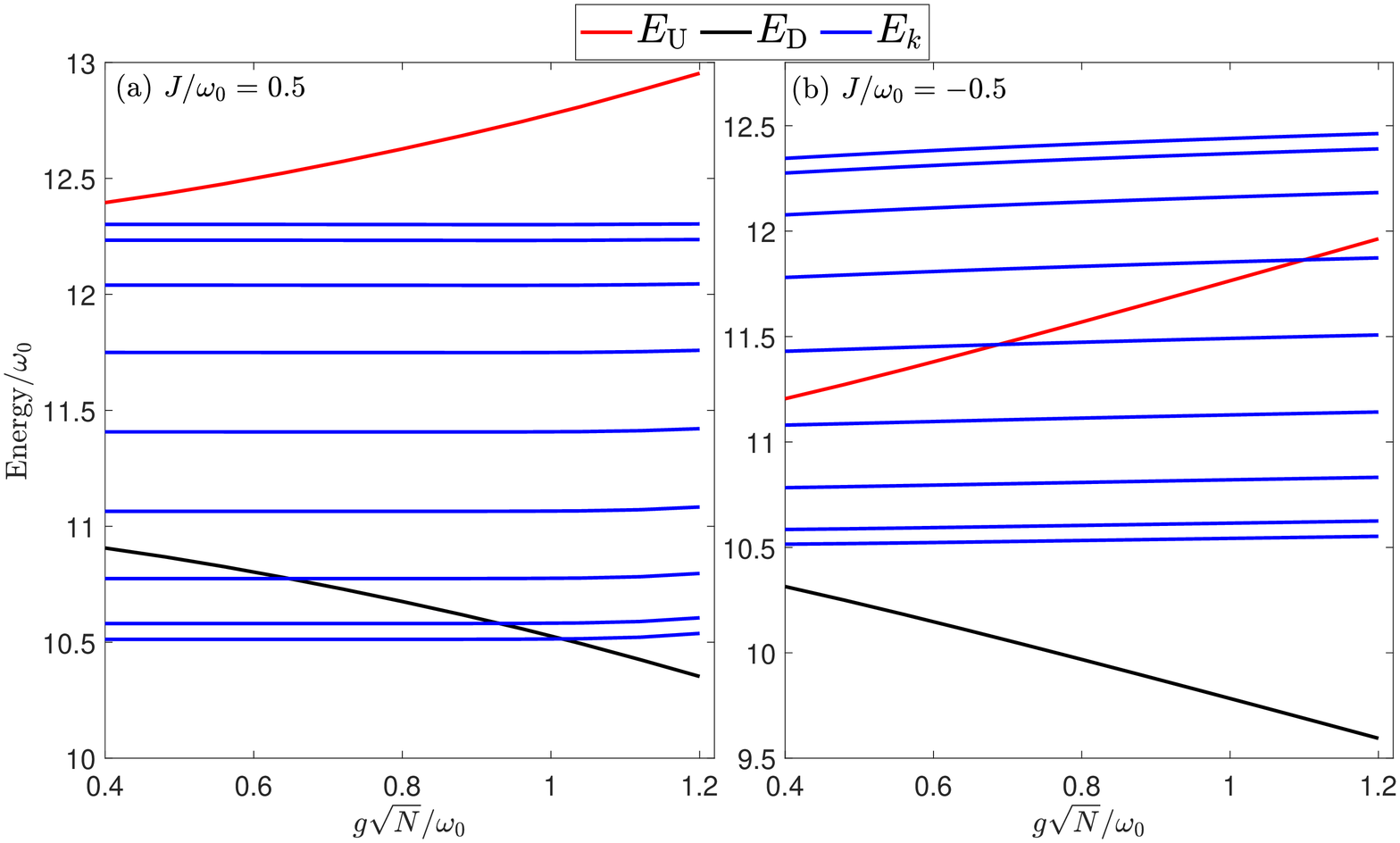}
\includegraphics[width=.52\textwidth]{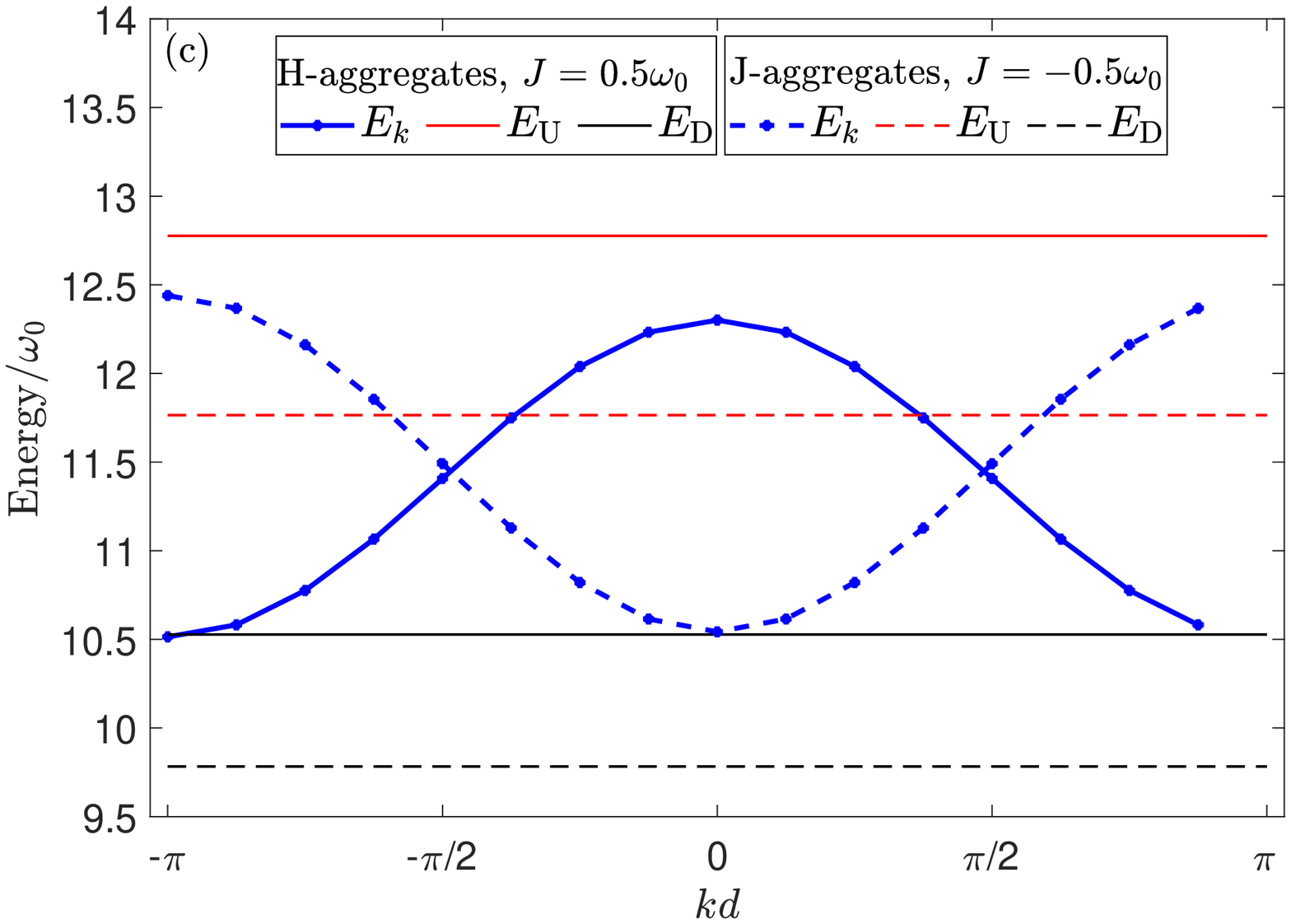}
\caption{Evolution of the energy levels with $g\sqrt{N}/\omega_0$ for (a) an H-aggregate with $J/\omega_0=0.5$; (b) a J-aggregate with $J/\omega_0=-0.5$ at a low temperature $1/\beta\omega_0=0.1$. The corresponding single-particle dispersions for $g\sqrt{N}/\omega_0=1$ are shown in (c). Other parameters: $N=16$, $\tau\omega_0=30$, $\varepsilon/\omega_0=11.8$, $\omega_{\mathrm{c}}/\omega_0=11$, and $\lambda=0.7$.}
\label{Fig4}
\end{figure}

 In Fig.~\ref{Fig3}(a) we present the reduced coherent mobility $\mu^{(\mathrm{coh})}/\mu_0$ for an H-aggregate as a function of temperature in the presence of the exciton-cavity coupling. Intriguingly enough, at relatively low temperatures with $k_BT<\omega_0$, we observe that the coherent mobility first increases with increasing exciton-cavity coupling and then decreases after passing a crossover cavity coupling at $g\sqrt{N}/\omega_0\sim 1$ [see the lower inset in Fig.~\ref{Fig3}(a)]. To understand this nonmonotonic behavior, we first note that only those Merrifield dark excitons with $\mathcal{E}_\eta$ close to $E_{\mathrm{U}}$ or $E_{\mathrm{D}}$ have significant contribution to the second term of Eq.~(\ref{mob-coh}). By investigating the evolution of the spectrum of an H-aggregate with increasing $g\sqrt{N}/\omega_0$ at a low temperature of $k_BT=0.1\omega_0$ [Fig.~\ref{Fig4}(a)], we see that the lower polaron polariton level $E_{\mathrm{D}}$ (black curve) experiences several intersections with the Merrifield dark exciton levels (blue curves) having decreasing energies, while the upper polaron polariton level $E_{\mathrm{U}}$ always separates from the Merrifield dark exciton band. This indicates that the lower polaron polariton state serves as a main transmission channel for an H-aggregate. In addition, the decline of the intersections leads to an enhancement in the coherent mobility due the increasing of the thermal factor $e^{-\beta \mathcal{E}_\eta}$. After passing the last intersection with the lowest Merrifield dark exciton level [with $k=-\pi$, see Fig.~\ref{Fig4}(c)] at $g\sqrt{N}/\omega_0\sim 1$, where a maximum of the coherent mobility is expected, $E_{\mathrm{D}}$ becomes separated from the Merrifield dark exciton band, resulting in a drop of the coherent mobility with further increasing of $g\sqrt{N}/\omega_0$.
\par
\begin{figure}
\includegraphics[width=.52\textwidth]{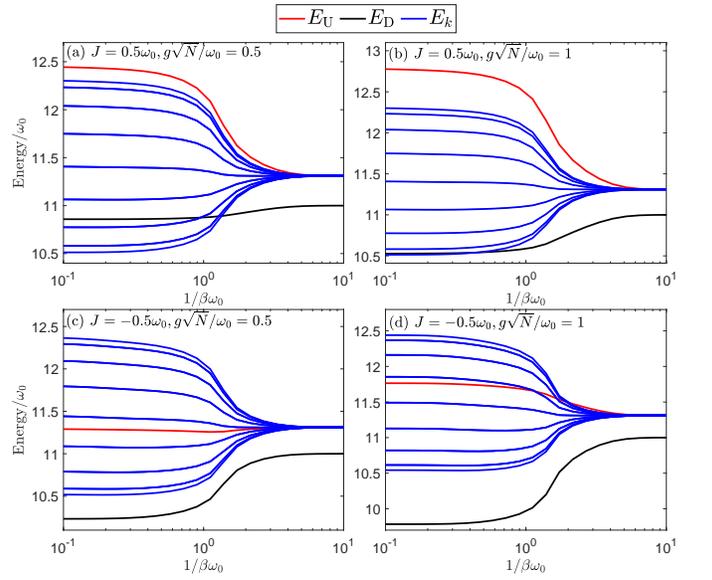}
\caption{Renormalized exciton spectrum as a function of temperature for various combinations of $(J/\omega_0, g\sqrt{N}/\omega_0)$. (a) $(J/\omega_0, g\sqrt{N}/\omega_0)=(0.5,0.5)$; (b) $(J/\omega_0, g\sqrt{N}/\omega_0)=(0.5,1)$; (c) $(J/\omega_0, g\sqrt{N}/\omega_0)=(-0.5,0.5)$; (d) $(J/\omega_0, g\sqrt{N}/\omega_0)=(-0.5,1)$. In all cases the Merrifield dark exciton bandwidth (blue curves) narrows down as temperature increases. Other parameters: $N=20$, $\varepsilon/\omega_0=11.8$, $\omega_{\mathrm{c}}/\omega_0=11$, and $\lambda=0.7$.}
\label{Fig5}
\end{figure}

 When $k_BT$ exceeds the vibrational energy $\omega_0$, the coherent mobility becomes largely insensitive to the variation of $g\sqrt{N}/\omega_0$. This is mainly owing to the band narrowing effect at high temperatures, which in turn narrows down the energy separations between neighboring Merrifield dark exciton levels [Fig.~\ref{Fig5}(a) and (b)].
\par A similar analysis can be applied to the case of a J-aggregate, for which $E_{\mathrm{D}}$ is separated from the Merrifield dark exciton band and the upper polaron polariton level is the main transfer channel [Fig.~\ref{Fig4}(b)]. In contrast to the case of an H-aggregate, the intersection between the upper polaron polariton and Merrifield dark exciton band moves up as $g\sqrt{N}/\omega_0$ increases, which explains the monotonic decreasing of the low-temperature coherent mobility for the J-aggregates shown in Fig.~\ref{Fig3}(b).
\par We also observe that after passing through a turning point at $k_BT\approx1.6\omega_0$ the mobility in both types of aggregates gets enhanced by the exciton-cavity coupling. From Fig.~\ref{Fig5} we see that in all cases considered the energy levels for the Merrifield dark excitons and the upper polaron polariton start to converge at the turning point due to the band narrowing effect, while the level of the lower polaron polariton is well separated from the continuous band. This indicates that the lower Merrifield polariton behaves more like a free photon in the high temperature regime, which is consistent with the fact that the weight of the Merrifield bright exciton, $S^2$, decreases as temperature increases [see the insets of Fig.~(\ref{Fig3})]. As a result, the second line in Eq.~(\ref{mob-coh}) will be dominated by the first term if $S^2$ is finite, which causes the increase in the mobility with increasing $g\sqrt{N}/\omega_0$ due to the prefactor $\sim (g\sqrt{N})^2$.
\subsection{Total mobility}
\par
\begin{figure}
\includegraphics[width=.48\textwidth]{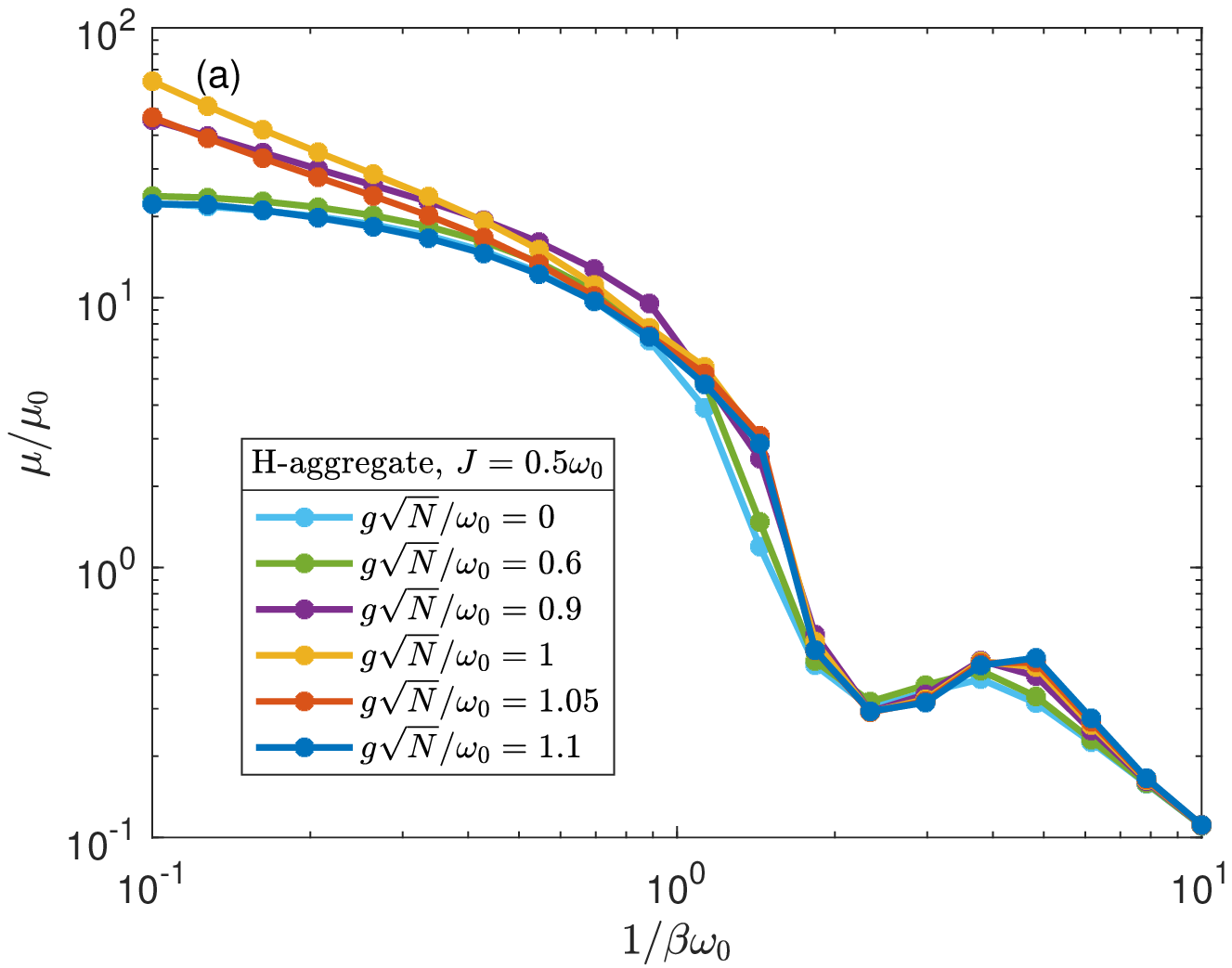}
\includegraphics[width=.48\textwidth]{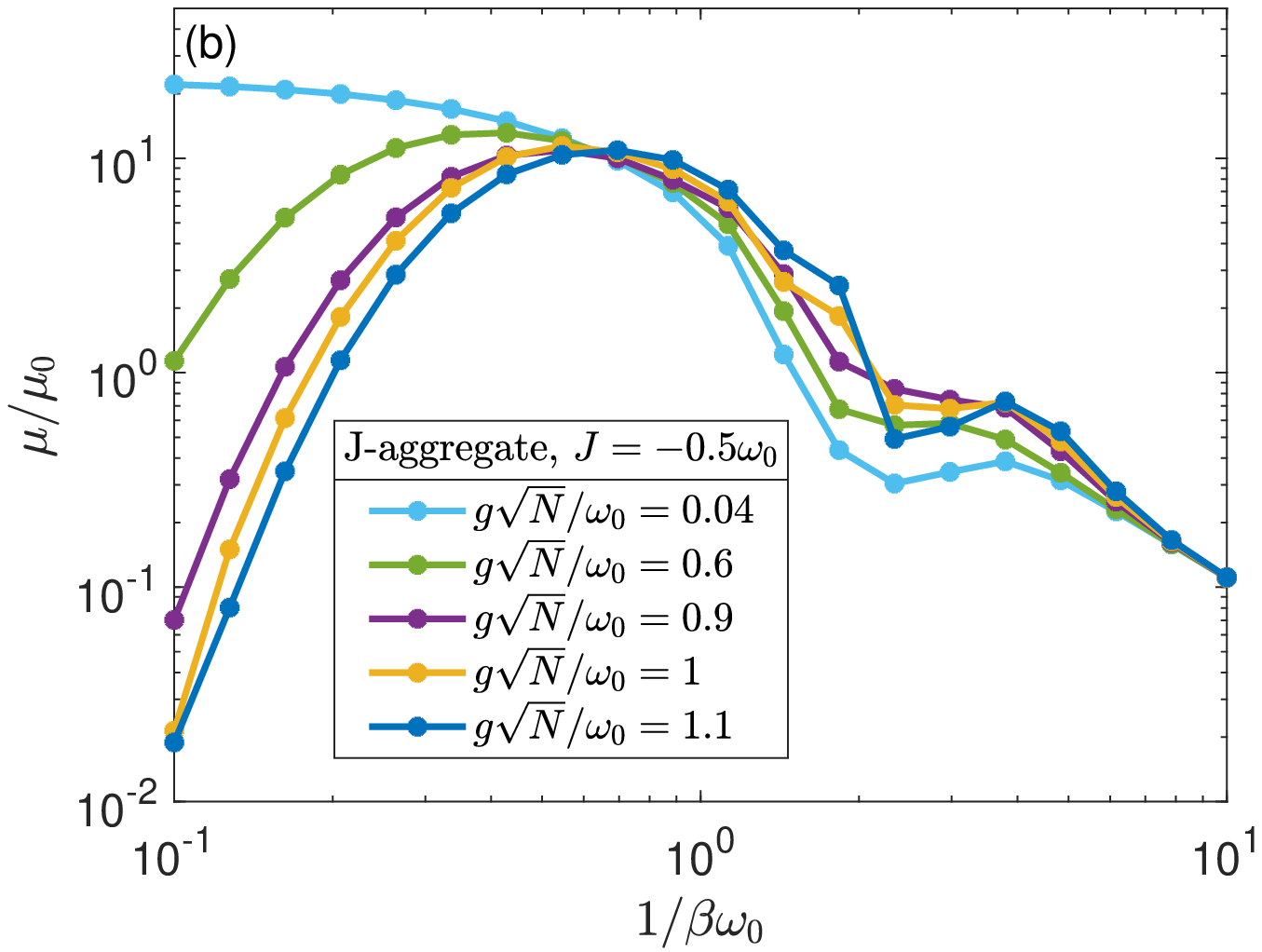}
\caption{Total exciton mobility of (a) an H-aggregate with $J/\omega_0=0.5$, (b) a J-aggregate with $J/\omega_0=-0.5$ for various values of $g\sqrt{N}/\omega_0$. Other parameters: $N=16$, $\tau\omega_0=30$, $\varepsilon/\omega_0=11.8$, $\omega_c/\omega_0=11$, and $\lambda=0.7$.}
\label{Fig6}
\end{figure}

 We now discuss the effect of exciton-cavity coupling on the total mobility of the molecular aggregates. Figure~\ref{Fig6}(a) [(b)] shows the total mobility as a function of temperature for the H-aggregate (J-aggregate).
At low temperatures, the evolution of the total mobility with increasing $g$ behaves similarly to that of the coherent mobility since the vibrational effects is minor.
Similar to the case of vanishing exciton-cavity coupling~\cite{Cheng2008,PRB2009}, a local maximum is observed in the high-temperature regime, which can be interpreted as the incoherent transport via polaron hopping.
For both the H- and J-aggregate, the total mobility exhibits a monotonic increase with increasing $g$ around the above-mentioned local maximum. This might be due to the cavity-induced enhancement of the vibrational dressing of the cavity mode (measured by the parameter $h$)~\cite{PRB2016}, which enters the thermal average of the vibrational operators appearing in the correlations $\langle\tilde{\mathcal{J}}_a(t)  \tilde{\mathcal{J}}_{\mathrm{c}} \rangle_{\tilde{H}}$, $\langle\tilde{\mathcal{J}}_{\mathrm{c}}(t)  \tilde{\mathcal{J}}_a \rangle_{\tilde{H}}$, and $\langle\tilde{\mathcal{J}}_{\mathrm{c}}(t)  \tilde{\mathcal{J}}_{\mathrm{c}} \rangle_{\tilde{H}}$.
\begin{figure}
\includegraphics[width=.48\textwidth]{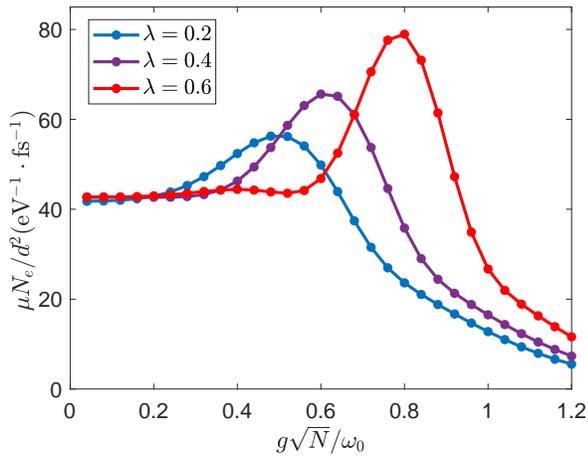}
\caption{The total mobility of an H-aggregate as a function of $g\sqrt{N}/\omega_0$ at temperature $T=298K$ and for $\lambda=0.2$, $0.4$, and $0.6$. Parameters: $N=16$, $J=0.5\omega_0$, $\omega_c=1.85$~eV, $\varepsilon_0=2$~eV, and $\omega_0=0.17$~eV, and $\tau=100$~fs.}
\label{Fig7}
\end{figure}
\par Similar to the coherent mobility, at low temperatures the total mobility also depend nonmonotonically on the exciton-cavity coupling.
\begin{figure}
\includegraphics[width=.48\textwidth]{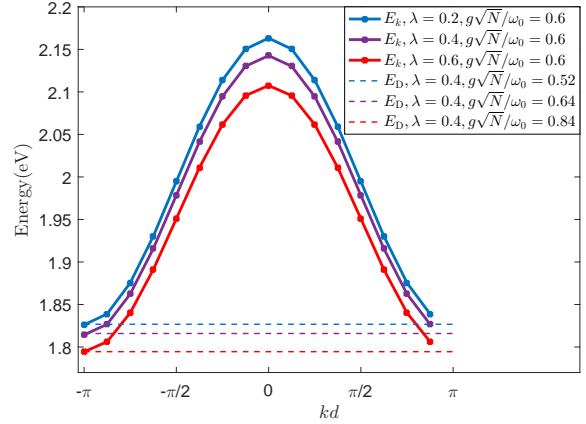}
\caption{ The solid curves show the dark exciton dispersion for $g\sqrt{N}/\omega_0=0.6$ and $\lambda=0.2$, $0.4$, and $0.6$. The three dashed lines represent the lower polaron polariton level $E_{\mathrm{D}}$ for a fixed $\lambda=0.4$ and $g\sqrt{N}/\omega_0=0.52$, $0.64$, and $0.84$. Other parameters are the same as those in Fig.~\ref{Fig7}.}
\label{Fig8}
\end{figure}
To illustrate this, we show in Fig.~\ref{Fig7} the total mobility as a function of $g\sqrt{N}/\omega_0$ at room temperature ($T=298$~K) for an H-aggregate with $J/\omega_0=0.5$. For each of the $\lambda$ considered, there always exists an optimal exciton-cavity coupling strength $g=g_{\mathrm{opt}}$ at which the mobility reaches a maximum. Moreover, both $g_{\mathrm{opt}}$ and the corresponding optimal mobility increase with increasing $\lambda$. To qualitatively understand these phenomena, let us look at the evolution of the dispersion when $\lambda$ and $g$ are varied. From numerical check we find that the energy levels of the dark excitons are insensitive to the change of $g$ for fixed $\lambda$ [see Fig.~\ref{Fig4}(b)], while the energy of the lower polaron polariton $E_{\mathrm{D}}$ is insensitive to the change of $\lambda$ for fixed $g$. We thus present in Fig.~(\ref{Fig8}) the dark exciton dispersions at a fixed exciton-cavity coupling $g\sqrt{N}/\omega_0=0.6$ and for $\lambda=0.2$, $0.4$, and $0.6$ (solid curves). We also plot the lower polaron polariton level $E_{\mathrm{D}}$ for a fixed $\lambda=0.4$ and $g\sqrt{N}/\omega_0=0.52$, $0.64$, and $0.84$ (dashed lines), in order to make the these energies consistent with the corresponding lowest dark exciton energies (with $k=-\pi$). It can be seen that the three values of $g\sqrt{N}/\omega_0$ roughly give the corresponding optimal exciton-cavity coupling $g_{\mathrm{opt}}$ shown in Fig.~(\ref{Fig7}). Actually, the system lies in the low temperature regime and the observed mobility can qualitatively be captured by coherent mobility given by Eq.~(\ref{mob-coh}). To understand the enhancement of the optimal mobility with increasing $\lambda$, we first note from Eq.~(\ref{mob-coh}) that the cavity contribution to the coherent mobility is proportional to $g^2_{\mathrm{opt}}$, which increases as $\lambda$ increases. We also note that the dark exciton band moves down as $\lambda$ increases, which results in a larger thermal occupation of the lowest dark exciton state.
\section{Conclusions}\label{V}
\par In this work, we present a microscopic theory for exciton transport in organic molecular crystals interacting with a single-mode cavity. Starting with the Holstein-Tavis-Cummings model, we employ a generalized Merrifield transformation developed in Ref.~\cite{PRB2016} to treat the system and obtained an expression for the exciton mobility based on the Kubo formula. As a generalization of the Cheng-Silbey~\cite{Cheng2008} method that deals with charge-carrier transport in molecular crystals without a cavity, our generalized variational canonical transformation not only takes into account the dressing of an exciton by neighboring vibrations, but also the vibrational dressing of the cavity mode. The method is believed to be capable of covering a wide range of parameters and temperatures.
\par Using the zeroth order of the exciton-vibration coupling, we derive a closed-form expression for the coherent contribution to the total mobility, which determines the behavior of the total mobility at low temperatures. Using the developed formalism, we perform numerical simulations on both the one-dimensional H- and J-aggregates. It is found that the exciton-cavity coupling can influence the transport properties in a significant way. Specifically, we find that for the H-aggregate there exists an optimal exciton-cavity coupling strength at which the total mobility is maximized, while for the J-aggregate the mobility decreases monotonically with increasing exciton-cavity coupling. However, an enhancement of the mobility is observed for both types of aggregates in the high temperature limit.\\

\noindent{\bf DATA AVAILABILITY}
\par The data that support the findings of this study are available from the corresponding author upon reasonable request.

\begin{acknowledgments}
We thank Dazhi Xu for useful discussions. This work was supported by the National Natural Science Foundation of China (NSFC) under Grant No. 11705007 and No. 11675014, and partially by the Beijing Institute of Technology Research Fund Program for Young Scholars.
\end{acknowledgments}

\appendix

\begin{widetext}
\section{Calculation of current-current correlation function Eq.~(\ref{cu-cu})}\label{AppA}
\par Using the approximated Hamiltonian $\tilde{H}\approx\tilde{H}_0$, we can derive analytical expressions for the four contributions in the  current-current correlation function given by Eq.~(\ref{cu-cu}).
\par 1) The exciton-exciton term $\langle\tilde{\mathcal{J}}_a(t)  \tilde{\mathcal{J}}_a \rangle_{\tilde{H}}$:
\begin{eqnarray}
\langle\tilde{\mathcal{J}}_a(t)  \tilde{\mathcal{J}}_a \rangle_{\tilde{H}}&=&-(dJ)^2\sum_{jj'}\langle a^\dag_{j}(t)a_{j+1}(t) a^\dag_{j'}a_{j'+1}\rangle_{\tilde{H}_{\mathrm{S}}}\langle e^{B_{j+1}(t)-B_j(t)}e^{B_{j'+1}-B_{j'}} \rangle_{H_{\mathrm{v}}}\nonumber\\
& &+(dJ)^2\sum_{jj'}\langle a^\dag_{j}(t)a_{j+1}(t) a^\dag_{j'+1}a_{j'}\rangle_{\tilde{H}_{\mathrm{S}}}\langle e^{B_{j+1}(t)-B_j(t)}e^{B_{j'}-B_{j'+1}} \rangle_{H_{\mathrm{v}}}\nonumber\\
& &+(dJ)^2\sum_{jj'}\langle a^\dag_{j+1}(t)a_{j}(t) a^\dag_{j'}a_{j'+1}\rangle_{\tilde{H}_{\mathrm{S}}}\langle e^{B_{j}(t)-B_{j+1}(t)}e^{B_{j'+1}-B_{j'}} \rangle_{H_{\mathrm{v}}}\nonumber\\
& &-(dJ)^2\sum_{jj'}\langle a^\dag_{j+1}(t)a_{j}(t) a^\dag_{j'+1}a_{j'}\rangle_{\tilde{H}_{\mathrm{S}}}\langle e^{B_{j}(t)-B_{j+1}(t)}e^{B_{j'}-B_{j'+1}} \rangle_{H_{\mathrm{v}}}.
\end{eqnarray}
We thus need to calculate
\begin{eqnarray}\label{aaaa}
&& \langle a^\dag_{r}(t)a_{l}(t) a^\dag_{m}a_{n}\rangle_{\tilde{H}_{\mathrm{S}}} =\frac{1}{N^2}\sum_{\eta\eta'\eta''\eta'''}e^{-iK_\eta rd+i\mathcal{E}_\eta t}x_\eta e^{iK_{\eta'} ld-i\mathcal{E}_{\eta'} t}x_{\eta'}e^{-iK_{\eta''} md}x_{\eta''} e^{iK_{\eta'''} nd }x_{\eta'''} \langle  f^\dag_\eta f_{\eta'} f^\dag_{\eta''} f_{\eta'''}\rangle_{\tilde{H}_{\mathrm{S}}},\nonumber\\
\end{eqnarray}
where we used $a_j(t)= \frac{1}{\sqrt{N}}\sum^{N+1}_{\eta=1} e^{iK_\eta jd-i\mathcal{E}_\eta t}x_\eta f_\eta$ and $c(t)=\sum^{N+1}_{\eta=1} e^{-i\mathcal{E}_\eta t}y_\eta f_\eta$. Recall that $f_{\eta}f^\dag_{\eta'}=|\mathrm{vac}\rangle\langle\eta |\eta'\rangle\langle \mathrm{vac}|=\delta_{\eta\eta'}|\mathrm{vac}\rangle\langle \mathrm{vac}|$, and $f^\dag_{\eta}f_{\eta'}=|\eta\rangle\langle \mathrm{vac}|\mathrm{vac}\rangle\langle \eta'|=|\eta\rangle \langle \eta'|$, so $f^\dag_\eta f_{\eta'} f^\dag_{\eta''} f_{\eta'''}=|\eta\rangle \langle \eta'| \eta''\rangle \langle \eta'''|=\delta_{\eta'\eta''}|\eta\rangle\langle\eta'''|$, yielding
\begin{eqnarray}\label{ffff}
\langle  f^\dag_\eta f_{\eta'} f^\dag_{\eta''} f_{\eta'''}\rangle_{\tilde{H}_{\mathrm{S}}}=\frac{1}{Z_{\mathrm{S}}}\delta_{\eta'\eta''}\mathrm{Tr_S}(e^{-\beta\tilde{H}_{\mathrm{S}}}|\eta\rangle\langle\eta'''|)=\frac{1}{Z_{\mathrm{S}}}\delta_{\eta'\eta''}\sum_{\chi}e^{-\beta \mathcal{E}_\chi}\langle\chi|\eta\rangle\langle\eta'''|\chi\rangle=\frac{e^{-\beta \mathcal{E}_\eta} }{Z_{\mathrm{S}}}\delta_{\eta'\eta''} \delta_{\eta\eta'''}.
\end{eqnarray}
Inserting Eq.~(\ref{ffff}) into Eq.~(\ref{aaaa}), we have
\begin{eqnarray}
&& \langle a^\dag_{r}(t)a_{l}(t) a^\dag_{m}a_{n}\rangle_{\tilde{H}_{\mathrm{S}}} =\frac{1}{Z_SN^2}\sum_{\eta  }e^{-\beta \mathcal{E}_\eta}x^2_\eta e^{-iK_\eta (r-n)d+i\mathcal{E}_\eta t}\sum_{\eta'} x^2_{\eta'} e^{iK_{\eta'} (l-m)d-i\mathcal{E}_{\eta'} t}.
\end{eqnarray}
Another quantity we need to calculate is the thermal average of the vibrational operators
\begin{eqnarray}\label{eBeBApp}
&&\langle e^{B_{l}(t)-B_r(t)}e^{B_{n}-B_{m}} \rangle_{H_{\mathrm{v}}} =e^{-\frac{1}{2}\Phi_{\omega_0}(0)\sum_s[(f_{s-l}-f_{s-r})^2+(f_{s-n}-f_{s-m})^2]}e^{-\Phi_{\omega_0}(t)\sum_s(f_{s-l}-f_{s-r})(f_{s-n}-f_{s-m})},
\end{eqnarray}
where $\Phi_{\omega_0}(t)=n_{\omega_0}e^{i\omega_0t}+(1+n_{\omega_0})e^{-i\omega_0t}$ with $n_{\omega_0}=1/(e^{\beta\omega_0}-1)$ the Bose-Einstein distribution function. Eq.~(\ref{eBeBApp}) can be derived in the standard way by using the Baker-Campbell-Hausdorff formula. From the relation $\Phi_{\omega_0}(0)=\coth\frac{\beta\omega_0}{2}$, it is readily seen that $e^{-\frac{1}{2}\Phi_{\omega_0}(0)\sum_s[(f_{s-l}-f_{s-r})^2+(f_{s-n}-f_{s-m})^2]}=\Theta_{|l-r|}\Theta_{|n-m|}$, where $\Theta_i$ is given by Eq.~(\ref{TT1}). So Eq.~\ref{eBeBApp} can be rewritten as
\begin{eqnarray}\label{eBeBApp1}
&&\langle e^{B_{l}(t)-B_r(t)}e^{B_{n}-B_{m}} \rangle_{H_{\mathrm{v}}} =\Theta_{|l-r|}\Theta_{|n-m|}e^{-\Phi_{\omega_0}(t)\sum_s(f_{s-l}-f_{s-r})(f_{s-n}-f_{s-m})}.
\end{eqnarray}
We will later use thermal averages involving $B_{\mathrm{c}}$, which can be obtained by replacing the corresponding $f$ by $h$ in the above equation, e.g.,
\begin{eqnarray}
\langle e^{B_{\mathrm{c}}(t)-B_r(t)}e^{B_{n}-B_{m}} \rangle_{H_v} &=&e^{-\frac{1}{2}\Phi_{\omega_0}(0)\sum_s[(h-f_{s-r})^2+(f_{s-n}-f_{s-m})^2]}e^{-\Phi_{\omega_0}(t)\sum_s[(h-f_{s-r})(f_{s-n}-f_{s-m})]}\nonumber\\
&=&\Theta\Theta_{|n-m|}e^{-\Phi_{\omega_0}(t)\sum_s[(h-f_{s-r})(f_{s-n}-f_{s-m})]}. \end{eqnarray}
\par 2) The cross term $\langle\tilde{\mathcal{J}}_a(t)  \tilde{\mathcal{J}}_{\mathrm{c}} \rangle_{\tilde{H}} $:
\begin{eqnarray}
 \langle\tilde{\mathcal{J}}_a(t)  \tilde{\mathcal{J}}_{\mathrm{c}} \rangle_{\tilde{H}} &=& dJg\sum_{jj'}R_{j'}\langle a^\dag_{j}(t)a_{j+1}(t) a^\dag_{j'}c\rangle_{\tilde{H}_{\mathrm{S}}}\langle e^{B_{j+1}(t)-B_j(t)}e^{B_{\mathrm{c}}-B_{j'}} \rangle_{H_{\mathrm{v}}}\nonumber\\
& &- dJg\sum_{jj'}R_{j'}\langle a^\dag_{j}(t)a_{j+1}(t) c^\dag a_{j'}\rangle_{\tilde{H}_{\mathrm{S}}}\langle e^{B_{j+1}(t)-B_j(t)}e^{B_{j'}-B_{\mathrm{c}}} \rangle_{H_{\mathrm{v}}}\nonumber\\
& &- dJg\sum_{jj'}R_{j'}\langle a^\dag_{j+1}(t)a_{j}(t) a^\dag_{j'}c\rangle_{\tilde{H}_{\mathrm{S}}}\langle e^{B_{j}(t)-B_{j+1}(t)}e^{B_{\mathrm{c}}-B_{j'}} \rangle_{H_{\mathrm{v}}}\nonumber\\
& &+ dJg\sum_{jj'}R_{j'}\langle a^\dag_{j+1}(t)a_{j}(t) c^\dag a_{j'}\rangle_{\tilde{H}_{\mathrm{S}}}\langle e^{B_{j}(t)-B_{j+1}(t)}e^{B_{j'}-B_{\mathrm{c}}} \rangle_{H_{\mathrm{v}}}.
\end{eqnarray}
The two types of the thermal averages of the exciton-photon operators are
\begin{eqnarray}
 \langle a^\dag_{r}(t)a_{l}(t) a^\dag_{m}c\rangle_{\tilde{H}_{\mathrm{S}}}&=&\frac{1}{Z_{\mathrm{S}} N\sqrt{N}}\sum_{\eta}e^{-\beta \mathcal{E}_\eta}x_\eta y_\eta e^{-iK_\eta rd+i\mathcal{E}_\eta t}\sum_{\eta'} x^2_{\eta'} e^{iK_{\eta'} (l-m)d-i\mathcal{E}_{\eta'} t},\nonumber\\
 \langle a^\dag_{r}(t)a_{l}(t) c^\dag a_{n}\rangle_{\tilde{H}_{\mathrm{S}}}&=&\frac{1}{Z_{\mathrm{S}} N\sqrt{N}}\sum_{\eta}e^{-\beta \mathcal{E}_\eta}x^2_\eta e^{-iK_\eta (r-n)d+i\mathcal{E}_\eta t}\sum_{\eta'} x_{\eta'}y_{\eta'} e^{iK_{\eta'} ld-i\mathcal{E}_{\eta'} t}.
\end{eqnarray}
\par 3) $\langle\tilde{\mathcal{J}}_{\mathrm{c}}(t)  \tilde{\mathcal{J}}_a \rangle_{\tilde{H}}$
\begin{eqnarray}
 \langle\tilde{\mathcal{J}}_{\mathrm{c}}(t)  \tilde{\mathcal{J}}_a \rangle_{\tilde{H}} &=& dJg\sum_{jj'}R_j\langle a^\dag_{j}(t)c(t) a^\dag_{j'}a_{j'+1}\rangle_{\tilde{H}_{\mathrm{S}}}\langle e^{B_{\mathrm{c}}(t)-B_j(t)}e^{B_{j'+1}-B_{j'}} \rangle_{H_{\mathrm{v}}}\nonumber\\
& &- dJg\sum_{jj'}R_j\langle a^\dag_{j}(t)c(t) a_{j'+1}^\dag a_{j'}\rangle_{\tilde{H}_{\mathrm{S}}}\langle e^{B_{\mathrm{c}}(t)-B_j(t)}e^{B_{j'}-B_{j'+1}} \rangle_{H_{\mathrm{v}}}\nonumber\\
& &- dJg\sum_{jj'}R_j\langle c^\dag(t)a_{j}(t) a^\dag_{j'}a_{j'+1}\rangle_{\tilde{H}_{\mathrm{S}}}\langle e^{B_{j}(t)-B_{\mathrm{c}}(t)}e^{B_{j'+1}-B_{j'}} \rangle_{H_{\mathrm{v}}}\nonumber\\
& &+ dJg\sum_{jj'}R_j\langle c^\dag(t)a_{j}(t) a_{j'+1}^\dag a_{j'}\rangle_{\tilde{H}_{\mathrm{S}}}\langle e^{B_{j}(t)-B_{\mathrm{c}}(t)}e^{B_{j'}-B_{j'+1}} \rangle_{H_{\mathrm{v}}},
\end{eqnarray}
with
\begin{eqnarray}
\langle c^\dag(t)a_{l}(t) a^\dag_{m}a_{n}\rangle_{\tilde{H}_{\mathrm{S}}}&=& \frac{1}{Z_{\mathrm{S}}N\sqrt{N}}\sum_{\eta}e^{-\beta \mathcal{E}_\eta}x_\eta y_{\eta} e^{iK_\eta nd+i\mathcal{E}_\eta t}\sum_{\eta'} x^2_{\eta'} e^{iK_{\eta'} (l-m)d-i\mathcal{E}_{\eta'} t}, \nonumber\\
 \langle a^\dag_{r}(t)c(t) a^\dag_{m}a_{n}\rangle_{\tilde{H}_{\mathrm{S}}}  &=&\frac{1}{Z_{\mathrm{S}}N\sqrt{N}}\sum_{\eta}e^{-\beta \mathcal{E}_\eta}x^2_\eta e^{-iK_\eta (r-n)d+i\mathcal{E}_\eta t}\sum_{\eta'} x_{\eta'}y_{\eta'} e^{-iK_{\eta'} md-i\mathcal{E}_{\eta'} t}.
\end{eqnarray}
\par 4) $\langle\tilde{\mathcal{J}}_{\mathrm{c}}(t)  \tilde{\mathcal{J}}_{\mathrm{c}} \rangle_{\tilde{H}}$
\begin{eqnarray}
 \langle\tilde{\mathcal{J}}_{\mathrm{c}}(t)  \tilde{\mathcal{J}}_{\mathrm{c}} \rangle_{\tilde{H}} &=&-g^2 \sum^N_{j=1}\sum^N_{j'=1} R_{j}R_{j'}\langle[a^\dag_{j}(t)c(t)e^{B_{\mathrm{c}}(t)-B_j(t)}-e^{B_{j}(t)-B_{\mathrm{c}}(t) }c^\dag(t)a_{j}(t)]
(a^\dag_{j'} c e^{B_c-B_{j'}}-e^{B_{j'}-B_c}c^\dag a_{j'}) \rangle_{\tilde{H}}\nonumber\\
&=&-g^2\sum_{jj'}R_jR_{j'}\langle a^\dag_{j}(t)c(t) a^\dag_{j'}c\rangle_{\tilde{H}_{\mathrm{S}}}\langle e^{B_{c}(t)-B_j(t)}e^{B_{c}-B_{j'}} \rangle_{H_{\mathrm{v}}}\nonumber\\
& &+g^2\sum_{jj'}R_jR_{j'}\langle a^\dag_{j}(t)c(t) c^\dag a_{j'}\rangle_{\tilde{H}_{\mathrm{S}}}\langle e^{B_{\mathrm{c}}(t)-B_j(t)}e^{B_{j'}-B_{\mathrm{c}}} \rangle_{H_{\mathrm{v}}}\nonumber\\
& &+g^2\sum_{jj'}R_jR_{j'}\langle c^\dag(t)a_{j}(t) a^\dag_{j'}c\rangle_{\tilde{H}_{\mathrm{S}}}\langle e^{B_{j}(t)-B_{\mathrm{c}}(t)}e^{B_{\mathrm{c}}-B_{j'}} \rangle_{H_{\mathrm{v}}}\nonumber\\
& &-g^2\sum_{jj'}R_jR_{j'}\langle c^\dag(t)a_{j}(t) c^\dag a_{j'}\rangle_{\tilde{H}_{\mathrm{S}}}\langle e^{B_{j}(t)-B_{\mathrm{c}}(t)}e^{B_{j'}-B_{\mathrm{c}}} \rangle_{H_{\mathrm{v}}},
\end{eqnarray}
where
\begin{eqnarray}
 \langle a^\dag_{r}(t)c(t) a^\dag_{m}c\rangle_{\tilde{H}_{\mathrm{S}}}  &=&\frac{1}{Z_{\mathrm{S}}N}\sum_{\eta}e^{-\beta \mathcal{E}_\eta}x_\eta y_\eta e^{-iK_\eta rd+i\mathcal{E}_\eta t}\sum_{\eta'} x_{\eta'}y_{\eta'} e^{-iK_{\eta'} md-i\mathcal{E}_{\eta'} t}, \nonumber\\
  \langle a^\dag_{r}(t)c(t) c^\dag a_{n}\rangle_{\tilde{H}_{\mathrm{S}}}  &=&\frac{1}{Z_{\mathrm{S}}N}\sum_{\eta}e^{-\beta \mathcal{E}_\eta}x^2_\eta e^{-iK_\eta (r-n)d+i\mathcal{E}_\eta t}\sum_{\eta'} y^2_{\eta'} e^{-i\mathcal{E}_{\eta'} t}, \nonumber\\
  \langle c^\dag(t)a_{l}(t) a^\dag_{m}c\rangle_{\tilde{H}_{\mathrm{S}}}  &=&\frac{1}{Z_{\mathrm{S}}N}\sum_{\eta}e^{-\beta \mathcal{E}_\eta}y^2_\eta e^{i\mathcal{E}_\eta t}\sum_{\eta'} x^2_{\eta'} e^{iK_{\eta'} (l-m)d-i\mathcal{E}_{\eta'} t}, \nonumber\\
   \langle c^\dag(t)a_{l}(t) c^\dag a_{n}\rangle_{\tilde{H}_{\mathrm{S}}}  &=&\frac{1}{Z_{\mathrm{S}}N}\sum_{\eta}e^{-\beta \mathcal{E}_\eta}x_\eta y_\eta e^{iK_\eta nd+i\mathcal{E}_\eta t}\sum_{\eta'} x_{\eta'} y_{\eta'} e^{iK_{\eta'} ld-i\mathcal{E}_{\eta'} t}.
\end{eqnarray}
\section{Calculation of the coherent mobility $\mu^{(\mathrm{coh})}$}\label{AppB}
\par We derive the explicit expression for the coherent contribution to the mobility, which is obtained by setting all the exponential factors in the thermal averages of the vibrational operators in Appendix~\ref{AppA} to be 1. The four terms are:
\par 1)
\begin{eqnarray}\label{JaJaApp}
 \langle\tilde{\mathcal{J}}_a(t)  \tilde{\mathcal{J}}_a \rangle^{(\mathrm{coh})}_{\tilde{H}} &=&-( d\tilde{J})^2\sum_{jj'}[\langle a^\dag_{j}(t)a_{j+1}(t) a^\dag_{j'}a_{j'+1}-\langle a^\dag_{j}(t)a_{j+1}(t) a^\dag_{j'+1}a_{j'}\rangle_{\tilde{H}_{\mathrm{S}}}\nonumber\\
& &-\langle a^\dag_{j+1}(t)a_{j}(t) a^\dag_{j'}a_{j'+1}\rangle_{\tilde{H}_{\mathrm{S}}}+\langle a^\dag_{j+1}(t)a_{j}(t) a^\dag_{j'+1}a_{j'}\rangle_{\tilde{H}_{\mathrm{S}}}]\nonumber\\
&=&-( d\tilde{J})^2 \frac{1}{Z_{\mathrm{S}} }\sum^{N-1}_{\eta=1 }e^{-\beta \mathcal{E}_\eta}  2 (\cos 2K_\eta d-1).
\end{eqnarray}
\par 2)
\begin{eqnarray}
 \langle\tilde{\mathcal{J}}_a(t)  \tilde{\mathcal{J}}_{\mathrm{c}} \rangle^{(\mathrm{coh})}_{\tilde{H}} &=& d\tilde{J}\tilde{g}\sum_{jj'}R_{j'}[\langle a^\dag_{j}(t)a_{j+1}(t) a^\dag_{j'}c\rangle_{\tilde{H}_{\mathrm{S}}}-\langle a^\dag_{j}(t)a_{j+1}(t) c^\dag a_{j'}\rangle_{\tilde{H}_{\mathrm{S}}}-\langle a^\dag_{j+1}(t)a_{j}(t) a^\dag_{j'}c\rangle_{\tilde{H}_{\mathrm{S}}}+\langle a^\dag_{j+1}(t)a_{j}(t) c^\dag a_{j'}\rangle_{\tilde{H}_{\mathrm{S}}}]\nonumber\\
&=&  d\tilde{J}\tilde{g}\sum_{ j'}R_{j'}\frac{1}{Z_{\mathrm{S}} \sqrt{N}}\sum_{\eta,\eta'=N,N+1}e^{-\beta \mathcal{E}_\eta}e^{i (\mathcal{E}_\eta-\mathcal{E}_{\eta'}) t} (x_\eta y_\eta  x^2_{\eta'}  - x_\eta y_\eta   x^2_{\eta'}  - x^2_\eta   x_{\eta'}y_{\eta'}  + x^2_\eta    x_{\eta'}y_{\eta'}  )\nonumber\\
&=&0.
\end{eqnarray}
\par 3) One can similarly show that $\langle\tilde{\mathcal{J}}_a(t)  \tilde{\mathcal{J}}_{\mathrm{c}} \rangle^{(\mathrm{coh})}_{\tilde{H}}=0$.
\par 4)
\begin{eqnarray}
 \langle\tilde{\mathcal{J}}_{\mathrm{c}}(t)  \tilde{\mathcal{J}}_{\mathrm{c}} \rangle_{\tilde{H}} &=&-\tilde{g}^2\sum_{jj'}R_jR_{j'}[\langle a^\dag_{j}(t)c(t) a^\dag_{j'}c\rangle_{\tilde{H}_{\mathrm{S}}}  -\langle a^\dag_{j}(t)c(t) c^\dag a_{j'}\rangle_{\tilde{H}_{\mathrm{S}}}-\langle c^\dag(t)a_{j}(t) a^\dag_{j'}c\rangle_{\tilde{H}_{\mathrm{S}}}+\langle c^\dag(t)a_{j}(t) c^\dag a_{j'}\rangle_{\tilde{H}_{\mathrm{S}}}]\nonumber\\
&=&-\tilde{g}^2\sum_{jj'}R_jR_{j'}\frac{1}{Z_{\mathrm{S}}N}\sum_{\eta\eta'}e^{-\beta\mathcal{E}_\eta}e^{i(\mathcal{E}_\eta-\mathcal{E}_{\eta'})t}\nonumber\\
&&[x_\eta y_\eta e^{-iK_\eta jd}x_{\eta'}y_{\eta'}e^{-iK_{\eta'}j'd}  -x^2_\eta  e^{-iK_\eta (j-j')d} y^2_{\eta'}  -  y^2_\eta  x^2_{\eta'} e^{-iK_{\eta'}(j-j')d} +x_\eta y_\eta e^{ iK_\eta j'd}x_{\eta'}y_{\eta'}e^{ iK_{\eta'}jd} ]\nonumber\\
&=& \tilde{g}^2\frac{1}{Z_{\mathrm{S}}N} [  e^{-\beta E_{\mathrm{U}}}e^{i(E_{\mathrm{U}}-E_{\mathrm{D}})t}  +e^{-\beta E_{\mathrm{D}}}e^{i(E_{\mathrm{D}}-E_{\mathrm{U}})t}    ]\sum_{jj'}R_jR_{j'} \nonumber\\
&&+\tilde{g}^2\frac{S^2}{Z_{\mathrm{S}}N}\sum^{N-1}_{\eta=1}[e^{-\beta\mathcal{E}_\eta}e^{i(\mathcal{E}_\eta-E_{\mathrm{U}})t}+e^{-\beta E_{\mathrm{U}}}e^{-i(\mathcal{E}_\eta-E_{\mathrm{U}})t}]   \sum_{jj'}R_jR_{j'} e^{-iK_\eta (j-j')d} \nonumber\\
&&+\tilde{g}^2\frac{ C^2}{Z_{\mathrm{S}}N}\sum^{N-1}_{\eta=1}[e^{-\beta\mathcal{E}_\eta}e^{i(\mathcal{E}_\eta-E_{\mathrm{D}})t}+e^{-\beta E_{\mathrm{D}}}e^{-i(\mathcal{E}_\eta-E_{\mathrm{D}})t}]  \sum_{jj'}R_jR_{j'}e^{-iK_\eta (j-j')d}.
\end{eqnarray}
By using $R_j=d(2j+1)/2$, we have
\begin{eqnarray}
\sum^{N/2-1}_{j=-N/2}R_j&=&d\sum^{N/2-1}_{j=-N/2}(2j+1)/2=0,\nonumber\\
\sum^{N/2-1}_{j=-N/2}R_je^{-iK_\eta jd}&=&dNe^{-\frac{1}{2}iK_\eta d(N-2)}\frac{1}{ 1-e^{iK_\eta d} },
\end{eqnarray}
so that
\begin{eqnarray}
\sum_{jj'}R_jR_{j'}&=&0,\nonumber\\
\sum_{jj'}R_jR_{j'}e^{-iK_\eta (j-j')d} &=& d^2N^2 \frac{ 1 }{(1-e^{iK_\eta d})(1-e^{-iK_\eta d})}=  \frac{ d^2N^2 }{2(1-\cos K_\eta d)},
\end{eqnarray}
giving
\begin{eqnarray}\label{JcJcApp}
 \langle\tilde{\mathcal{J}}_{\mathrm{c}}(t)  \tilde{\mathcal{J}}_{\mathrm{c}} \rangle_{\tilde{H}} &=&
 ( d\tilde{g})^2\frac{NS^2}{2Z_{\mathrm{S}} }\sum^{N-1}_{\eta=1}[e^{-\beta\mathcal{E}_\eta}e^{i(\mathcal{E}_\eta-E_{\mathrm{U}})t}+e^{-\beta E_{\mathrm{U}}}e^{-i(\mathcal{E}_\eta-E_{\mathrm{U}})t}]     \frac{ 1 }{ 1-\cos K_\eta d } \nonumber\\
&&+( d\tilde{g})^2\frac{ NC^2}{2Z_{\mathrm{S}} }\sum^{N-1}_{\eta=1}[e^{-\beta\mathcal{E}_\eta}e^{i(\mathcal{E}_\eta-E_{\mathrm{D}})t}+e^{-\beta E_{\mathrm{D}}}e^{-i(\mathcal{E}_\eta-E_{\mathrm{D}})t}]    \frac{ 1 }{ 1-\cos K_\eta d }.
\end{eqnarray}
By inserting Eqs.~(\ref{JaJaApp}) and (\ref{JcJcApp}) into Eq.~(\ref{mob}) we finally get Eq.~(\ref{mucoh}).
\end{widetext}

\end{document}